\documentclass[5p,frozencache,times]{elsarticle} 
\usepackage[colorlinks]{hyperref}

\usepackage{amssymb}

\usepackage[T1]{fontenc} 
\usepackage{lmodern} 
\usepackage[final]{microtype} 
\usepackage{amsmath} 
\usepackage{booktabs} 
\usepackage{url} 
\usepackage{alltt} 
\usepackage{subfig} 
\usepackage[linesnumbered,ruled,vlined]{algorithm2e} 
\usepackage{rotating} 
\usepackage{xcolor} 
\usepackage{minted} 

\usepackage{multirow}

\usepackage{tikz}
\usetikzlibrary{shapes.geometric, arrows.meta, fit, positioning, backgrounds, decorations.pathreplacing}

\usepackage{listings}
\newcommand{\CC}{C\nolinebreak\hspace{-.05em}\raisebox{.17ex}{\footnotesize\bf +}\nolinebreak\hspace{-.2em}\raisebox{.17ex}{\footnotesize\bf +}}

\newcommand{\pmf}[1]{#1}   
\newcommand{\ab}[1]{#1}  

\graphicspath{{figures/}{./}}

\journal{Future Generation Computer Systems}

\begin{document}

\begin{frontmatter}



\title{A Task-based Data-flow Methodology for Programming Heterogeneous Systems with Multiple Accelerator APIs\tnoteref{copyright}}
\tnotetext[copyright]{
Published in Future Generation Computer Systems Volume 180, July 2026, 108383. doi:\href{https://doi.org/10.1016/j.future.2026.108383}{10.1016/j.future.2026.108383} \\[0.5em]
\textcopyright 2026. This manuscript version is made available under the CC BY-NC-ND 4.0 license {\url{https://creativecommons.org/licenses/by-nc-nd/4.0/}}
}



\author[label1]{Aleix Boné\corref{cor1}}
\ead{aleix.boneribo@bsc.es}

\author[label1]{Alejandro Aguirre\corref{cor1}}
\ead{alejandro.aguirre@bsc.es}

\author[label1]{David Álvarez}
\ead{david.alvarez@bsc.es}

\author[label2,label1]{Pedro J.~Martinez-Ferrer}
\ead{pedro.martinez.ferrer@upc.edu}

\affiliation[label1]{organization={Barcelona Supercomputing Center~(BSC)},
            addressline={Pl.~Eusebi G\"uell~\mbox{1-3}},
            city={Barcelona},
            postcode={08034},
            state={Catalu\~{n}a},
            country={Spain}}

\affiliation[label2]{organization={Departament d'Arquitectura
            de Computadors~(DAC), Universitat Polit\`ecnica de Catalunya -
            BarcelonaTech~(UPC)},
            addressline={\\Campus~Nord, Edif.~D6, C.~Jordi~Girona~\mbox{31}},
            city={Barcelona},
            postcode={08034},
            state={Catalu\~{n}a},
            country={Spain}}

\author[label1]{Vicen\c{c} Beltran}
\ead{vbeltran@bsc.es}

\cortext[cor1]{Corresponding author}

\begin{abstract}
Heterogeneous nodes that combine multi-core CPUs with diverse accelerators are rapidly becoming the norm in both high-performance computing (HPC) and AI infrastructures.
Exploiting these platforms, however, requires orchestrating several low-level accelerator APIs such as CUDA, SYCL, and Triton. In some occasions they can be combined with optimized vendor math libraries: e.g., cuBLAS and oneAPI.  Each API or library introduces its own abstractions, execution semantics, and synchronization mechanisms.
Combining them within a single application is therefore error-prone and labor-intensive. We propose reusing a task-based data-flow methodology together with Task-Aware APIs (TA-libs) to overcome these limitations and facilitate the seamless integration of multiple accelerator programming models, while still leveraging the best-in-class kernels offered by each API.

Applications are expressed as a directed acyclic graph (DAG) of host tasks and device kernels managed by an OpenMP/OmpSs-2 runtime. We introduce Task-Aware SYCL (TASYCL) and leverage Task-Aware CUDA (TACUDA), which elevate individual accelerator invocations to first-class tasks. When multiple native runtimes coexist on the same multi-core CPU, they contend for threads, leading to oversubscription and performance variability. To address this, we unify their thread management under the nOS-V tasking and threading library, to which we contribute a new port of the P{o}CL (Portable OpenCL) runtime.

The methodology is evaluated on a multi-core server and a GPU-accelerated node using two contrasting workloads: the GPT-2 pre-training phase, representative of modern AI pipelines, and the HPCCG conjugate-gradient benchmark, representative of traditional HPC. From a performance standpoint, monolithic-kernel and fork-join executions are comparable ---in both execution time and memory footprint--- to a coarse-grained task-based formulation on both GPU-accelerated and multi-core systems. On the latter, unifying all runtimes through nOS-V mitigates interference and delivers performance on par with using a single runtime in isolation.

These results demonstrate that task-aware libraries, coupled with the nOS-V library, enable a single application to harness multiple accelerator programming models transparently and efficiently. The proposed methodology is immediately applicable to current heterogeneous nodes and is readily extensible to future systems that integrate even richer combinations of CPUs, GPUs, FPGAs, and AI accelerators.
\end{abstract}




\end{frontmatter}


\section{Introduction}\label{sec:introduction}
Traditionally, early HPC systems have relied on single-core processors with steadily increasing clock frequencies to improve performance; however, as frequency scaling became limited by power consumption and thermal constraints, multi-core architectures emerged as the new performance paradigm. Further advancements have led to heterogeneous systems incorporating specialized accelerators, notably GPUs, offering significantly higher efficiency for specific computational tasks at the cost of increased programming complexity and reduced flexibility. Recently, the rapid evolution of high-performance computing (HPC) and artificial intelligence (AI) workloads~\cite{Yi2019, Huerta2020} has significantly impacted the hardware and software landscapes, imposing complex demands on parallel programming models. 

The OpenMP~\cite{Dagum1998-OpenMP} application programming interface was originally designed to exploit multi-core architectures by annotating sequential loops using a fork-join execution model. OpenMP has evolved over time incorporating task parallelism and data-flow execution~\cite{Duran2008, Duran2009} to better manage irregular and unbalanced computations on multi-core platforms.

The introduction of CUDA~\cite{CUDA} in 2007 marked a significant milestone by effectively harnessing GPU parallelism for general-purpose computing beyond graphics-specific domains. While highly successful and broadly adopted, CUDA remains primarily specialized in orchestrating massively parallel GPU kernels coordinated from host multi-core CPUs.

The trend towards increasingly heterogeneous platforms has forced applications to shift from monolithic kernels, once favored due to expensive host-device data transfers, towards sophisticated multi-kernel applications with complex interaction~\cite{HeteroBench}. Modern systems frequently integrate CPUs and accelerators interconnected through high-bandwidth, low-latency interfaces, often sharing virtual or even physical memory hierarchies~\cite{Roberts2018, GraceHopper}. This shift, combined with the growing intersection of HPC and AI workloads, demands precise orchestration and synchronization of multiple activities, including host tasks, device kernels, and I/O operations, which in turn requires the combination of multiple APIs within the same application.

Numerous programming models have emerged after CUDA to manage the complexities introduced by increasingly diverse accelerators. OpenCL~\cite{OpenCL} provides a vendor-neutral, low-level alternative compatible across various accelerators, which has been later complemented by SYCL~\cite{SYCL}, a high-level and portable \CC{} standard supporting accelerated computing. Meanwhile, OpenACC~\cite{OpenACC} offers a simplified pathway towards accelerated computing by annotating existing sequential code, similar to the recent OpenMP offloading extensions. The rapid adoption of AI-specific workloads has also introduced specialized models like Triton~\cite{Triton}, optimized for data locality and reuse through tiled computations, supporting various backend devices including GPUs and multi-core CPUs.

Despite these advancements, generating highly optimized kernels for sophisticated hardware remains a challenging task, typically addressed through vendor-specific, highly-tuned libraries: cuBLAS, cuSPARSE, cu\-DNN, one\-MKL \ldots This complexity is further amplified by contemporary HPC systems increasingly incorporating multiple accelerators, ranging from discrete GPUs to integrated AI accelerators. This trend is accelerating as systems evolve from configurations with a single accelerator type (typically a GPU) to more heterogeneous designs that include GPUs, FPGAs, and integrated tensor cores on multi-core CPUs. As a result, applications will increasingly require the coordinated use of multiple APIs ---each targeting a different accelerator--- to fully exploit the capabilities of modern and future systems.

This integration presents significant challenges. First, combining multiple accelerator-specific APIs such as CUDA, ROCm~\cite{ROCm}, SYCL, or Triton at the programming level is inherently difficult. Each API introduces its own abstractions, execution models, and synchronization primitives, which are often incompatible or conceptually misaligned. This complicates the coordination of data movement, execution ordering, and inter-operation across different devices. Second, these APIs are typically backed by runtime systems that assume exclusive control over their respective hardware resources. When multiple runtimes coexist on the same node, each trying to manage shared multi-core CPUs or accelerators independently, resource contention and suboptimal sharing can lead to performance degradation or unpredictable behavior.


In this context, we introduce task-aware libraries, designed to seamlessly integrate \ab{multiple} APIs with task-based programming models. Leveraging the powerful and flexible data-flow execution model provided by OpenMP and OmpSs-2~\cite{Perez2017-OmpSs-2}, applications are represented as directed acyclic graphs (DAGs), with each node representing either host tasks or accelerator kernels. This abstraction simplifies synchronization and orchestration across diverse accelerators and runtime environments. Kernels are programmed using the most appropriate paradigm: i) SYCL for portable acceleration, ii) Triton for optimized AI computations, iii) OpenMP target offload for incrementally porting sequential kernels, or iv) native vendor libraries like cuBLAS/cuSPARSE and oneMKL for performance-critical tasks. Meanwhile, the OmpSs-2 or OpenMP \ab{runtimes} automatically \ab{manage} synchronization through DAG edges derived from data dependencies.
 
This paper contributes to a unified methodology that enables the seamless integration of multiple programming models (including SYCL, Triton, and OpenMP offload) alongside optimized vendor libraries such as cuBLAS/cu\-SPARSE and oneMKL, within a single coherent application. Building upon our previous work on task-aware libraries for message-passing, such as
TAMPI and TAGASPI~\cite{Sala2021-combining}\ab{, as well as TACUDA~\cite{Sala2024-ALPI}, a task-aware library for CUDA that
enables transparent and efficient integration of CUDA accelerator API with task-based programming models.
We now introduce Task-Aware SYCL (TASYCL),
a task-aware library which allows SYCL kernels to be embedded as tasks within a data-flow DAG, enabling the runtime to manage dependencies, synchronize execution, and orchestrate resources automatically.
\pmf{Although our implementation focuses on SYCL's specific API, it does follow} the same methodology as TACUDA, which reflects that
the underlying methodology initially presented in TACUDA is general and naturally extensible to other accelerator programming environments. This flexibility positions our approach as a portable and extensible solution for harnessing the growing diversity of accelerators in current and next-generation HPC systems.}

An additional challenge when combining multiple programming models within the same application is the interference between their respective runtime systems, particularly on the CPU side.  OpenMP, OmpSs-2, SYCL or OpenCL implementations like PoCL runtimes typically assume full control over core allocation, thread scheduling, and task management, which can lead to contention, oversubscription, and unpredictable performance when co-executed. In this work, we analyze these runtime interferences and propose the nOS-V tasking and threading library \ab{\cite{alvareznosv}} to mitigate them. By unifying the underlying execution model, nOS-V enables coordinated scheduling while mitigating interferences between different runtimes. To this end, we leverage the existing NODES/OmpSs-2 and libompv/OpenMP runtimes, which are already ported to nOS-V, and additionally contribute a new port of the P\ab{o}CL runtime to nOS-V, enabling OpenCL and SYCL workloads to benefit from the same unified execution layer. 

We evaluate our approach on two hardware platforms: a multi-core CPU system and a GPU-accelerated system. On the multi-core platform, we compare a traditional fork-join execution against a task-based data-flow execution to assess the benefits of fine-grained dependency management. On the GPU-based platform, we compare a monolithic kernel execution with our proposed task-based data-flow approach that integrates diverse accelerator APIs. For both platforms, we use two representative workloads: the GPT-2 pre-training phase~\cite{Karpathy2024-llm.c}, exemplifying an AI-centric application, and the high-performance conjugate gradient (HPCCG) benchmark~\cite{Heroux2017-HPCCG}, representing a traditional HPC workload. These experiments demonstrate the feasibility and effectiveness of integrating multiple programming models and runtimes within a single application, while highlighting the advantages of task-based orchestration across heterogeneous architectures.

The remainder of this paper is organized as follows. Section~\ref{sec:background} provides background on task-based programming models, accelerator APIs, and runtime systems. Section~\ref{sec:related} discusses related work on interoperability and task-aware execution in heterogeneous systems. Section~\ref{talibs} presents our proposed methodology and describes the design of task-aware libraries for integrating diverse accelerator APIs. Section~\ref{sec:evaluation} details our experimental evaluation using representative HPC and AI workloads. Finally, Section~\ref{sec:conclusions} concludes the paper and outlines directions for future work.

\section{Background}
\label{sec:background}
In this section, we begin by summarizing the main vendor-specific and portable APIs used to program modern heterogeneous systems, as well as the key libraries built on top of them. We then introduce task-based runtime systems, focusing on OmpSs-2 and OpenMP, and discuss the role of task-aware libraries in coordinating computation and data movement across CPUs and accelerators. Finally, we describe the two application benchmarks, GPT-2 pre-training and HPCCG and the parallelization strategies used in our study.

\subsection{Vendor APIs and libraries}
Vendor-specific APIs provide low-level access to accelerators, typically comprising two main components: i) a runtime or driver API for managing devices, memory allocation, data transfers, and host-device synchronization; and ii) a kernel programming language for writing compute kernels that exploit the accelerator's architectural features. In this work, we focus on APIs from the three main vendors (NVIDIA, AMD, and Intel) as representative examples. However, the proposed methodology is not restricted to these APIs and can be readily extended to support other accelerator-specific interfaces as well.

NVIDIA's CUDA provides both a runtime API for device management, memory handling, and kernel launches, as well as a lower-level driver API that offers fine-grained control. CUDA kernels are written in CUDA \CC{}, an extension of \CC{}, enabling efficient mapping of parallel algorithms onto the massively parallel architecture of NVIDIA GPUs. AMD's ROCm exposes the HIP API, which is also \CC{}-based, enabling code migration with minimal changes. Intel's Level Zero is a low-level interface intended for explicit control over Intel GPUs, providing fine-grained management of devices, memory, and command queues, and acting as the backend for higher-level abstractions such as SYCL. Although these APIs share similar concepts (e.g., streams in CUDA \ab{and HIP, or command} queues in Level Zero) they are not interoperable; for example, a CUDA stream cannot be directly used within a SYCL queue. This calls for careful coordination when integrating multiple APIs within the same application code.

All major vendors also provide highly optimized libraries to accelerate common computational kernels. NVIDIA offers cuBLAS/cuSPARSE, cuDNN, and cuFFT for linear algebra, deep learning, and Fourier transforms, respectively. AMD provides rocBLAS and rocFFT, as well as HIP interfaces like hipBLAS and hipFFT that mirror their CUDA counterparts. Intel supplies oneMKL, a math library that presents a unified SYCL interface and internally dispatches to optimized backend implementations, including those from NVIDIA and AMD. While these libraries enable high performance standard operations, they are typically confined to their respective ecosystems, which limits interoperability in heterogeneous environments.


\subsection{Portable APIs and libraries}
Portable APIs and libraries are designed to provide a unified programming interface across different hardware vendors and platforms, allowing developers to write applications that can target a wide range of devices with minimal code changes. The most widely adopted portable APIs include OpenCL, SYCL, OpenACC, OpenMP offload, and Triton, each with distinct approaches and design goals.

OpenCL offers a low-level, C-based runtime and kernel programming model that supports a broad set of devices, from CPUs and GPUs to FPGAs, through a single API. SYCL builds on top of OpenCL and provides a modern \CC{} abstraction layer, supporting single-source programming where host and device codes coexist in the same file. SYCL's concepts of queues, buffers, and kernel submission are closely aligned with those of CUDA and HIP, making it familiar to users of vendor-specific APIs while enabling portability across hardware from multiple vendors.

In contrast, OpenACC and OpenMP offload provide a pragma-based approach for heterogeneous programming. Rather than writing explicit kernel functions and memory management code, developers annotate sequential code with directives (\texttt{\#pragma acc} or \texttt{\#pragma omp target}) to specify data movement, parallelism, and device execution. The compiler and runtime generate the necessary device code and manage host-device synchronization, greatly simplifying incremental porting and maintenance of existing applications.

Triton takes a different approach, offering a Python-based domain-specific language for writing highly efficient GPU kernels, particularly for AI and deep learning workloads. While Triton generates device code that can run on different vendor accelerators, it does not provide a runtime API for host-device management; instead, it relies on the underlying vendor APIs (CUDA or ROCm) for memory management and synchronization.

Several portable libraries have emerged to complement these APIs. For example, clBLAS and CLBlast provide portable linear algebra routines for OpenCL, while oneMKL offers SYCL-based interfaces that can dispatched to optimized vendor libraries. These libraries, together with the APIs, form the foundation for developing portable and efficient applications for modern heterogeneous systems.

\subsection{OmpSs-2 and OpenMP task-based data-flow model}
The task-based data-flow model provides a flexible and expressive framework for developing parallel programs. In this approach, the computation is structured as a set of tasks, each representing a unit of work such as computation, data movement, or communication. Tasks specify their data dependencies (inputs, outputs, and input/output parameters) allowing the runtime system to represent the application as a directed acyclic graph (DAG). In this graph, nodes correspond to tasks and edges encode dependencies arising from data usage.

Both OmpSs-2 and OpenMP support this model through task constructs and explicit dependency clauses. Programmers define tasks using pragmas or directives and annotate them with dependency information. At runtime, the system analyzes these dependencies, builds the DAG dynamically, and schedules tasks as soon as their input data is ready, enabling efficient overlap of computation, communication, and I/O. This allows for fine-grained parallelism and effective utilization of available resources, while abstracting away low-level thread management.

This model can naturally capture a wide range of application patterns, from regular computations to irregular or data-driven workflows. Data-flow runtimes such as StarPU~\cite{Augonnet2011-StarPU}, PaRSEC~\cite{Bosilca2013-PaRSEC}, and OmpSs-2~\cite{Perez2017-OmpSs-2} extend this approach, supporting heterogeneous architectures and automatic resource management across CPUs, GPUs, and other accelerators. Expressing computation, communication, and I/O as a data-flow DAG allows programmers to easily manage complex interactions and maximize concurrency, making this model well-suited for modern heterogeneous systems.

\subsection{Task-aware libraries}
Task-Aware libraries (TA-libs) are designed to bridge the gap between high-performance APIs, including message-passing or offloading APIs, and task-based programming models. The key idea is to enable seamless and efficient integration of operations such as communication, I/O, or kernel offload within task-based runtimes like OmpSs-2 and OpenMP, while preserving correctness and maximizing concurrency.

Representative TA-libs include TAMPI (Task-Aware MPI) and TAGASPI (Task-Aware GASPI)~\cite{Sala2021-combining}. These libraries provide interfaces that allow blocking and non-blocking operations (MPI and GASPI communication calls) to be safely invoked within user-defined tasks. Internally, they leverage runtime support to pause and resume tasks, manage dependencies, and handle completion notifications. This enables overlapping computation and communication without deadlocks or unnecessary serialization, thus exposing more parallelism to the runtime scheduler.

To ensure portability and extensibility, TA-libs can be integrated with multiple runtimes via the ALPI interface~\cite{Sala2024-ALPI}. This allows OmpSs-2, OpenMP, and potentially other task-based systems to benefit from task-aware interfaces with minimal changes to user code or application logic.

It is important to distinguish TA-libs from directory/cache (D/C) mechanisms implemented by advanced data-flow runtimes such as OmpSs-2, StarPU~\cite{Augonnet2011-StarPU}, KAAPI~\cite{Gautier2007-KAAPI}, and IRIS~\cite{Kim2021-IRIS}, among others. D/C refers to the runtime's ability to automatically and transparently manage data movement between host and devices based solely on the input and output annotations specified on tasks. This means that the user does not need to insert explicit API calls to perform data transfers; instead, the runtime analyzes task dependencies, determines the necessary data movement, and automatically ensures that the required data is available in the appropriate memory space (host or device) before executing a kernel or task. This transparency gives the runtime maximum flexibility to schedule tasks on different resources and can simplify programming, as the burden of managing memory coherence and transfers is removed from the user.

On the other hand, task-aware libraries provide a more explicit programming model, giving developers fine-grained control over where and how tasks and operations are executed. By directly using the underlying accelerator or communication APIs, programmers can fully exploit hardware features and advanced API capabilities that may not be accessible through the generic D/C model. This approach can be particularly beneficial when applications require specialized synchronization, custom memory management, or advanced features not supported by transparent D/C mechanisms.

In practice, task-aware libraries and D/C models are complementary: TA-libs can be used within D/C frameworks to extend support for complex APIs, blocking operations, or fine-tuned performance scenarios where explicit control is desirable. In this paper, we focus on the use of task-aware libraries as a practical and portable solution for integrating multiple accelerator APIs and communication libraries within task-based data-flow applications.

\ab{
\subsection{nOS-V unified scheduling}

The nOS-V~\cite{alvareznosv} tasking library provides a unified threading
substrate with a centralized scheduler. Traditional oversubscription suffers
from problems like Lock Holder Preemption (LHP) and scalability collapse because
multiple applications independently manage resources through separate runtime
instances. In contrast, nOS-V maintains a single shared scheduler that
coordinates tasks from all applications. This ensures only one task runs per
physical core at any time, eliminating oversubscription issues. Tasks
voluntarily yield control rather than being forcibly preempted.
}

\subsection{GPT-2 pre-training}%
\label{sec:background:gpt2}
The GPT-2 model was originally presented in~\cite{Radford2019-GPT2}
where the authors demonstrated that as language models became larger,
and were trained on vast amounts of datasets (e.g., WebText with over
8$\,$M documents and 40$\,$GB of text), their ability to learn language
processing tasks without the need of explicit supervision emerged
naturally~\cite{Wei2022-emergentLLMs}.  GPT-2 was built on top of its
predecessor (i.e., GPT~\cite{Radford2018-GPT}) and both language
models shared the popular Transformer architecture~\cite{Vaswani2017-attTransformer}.

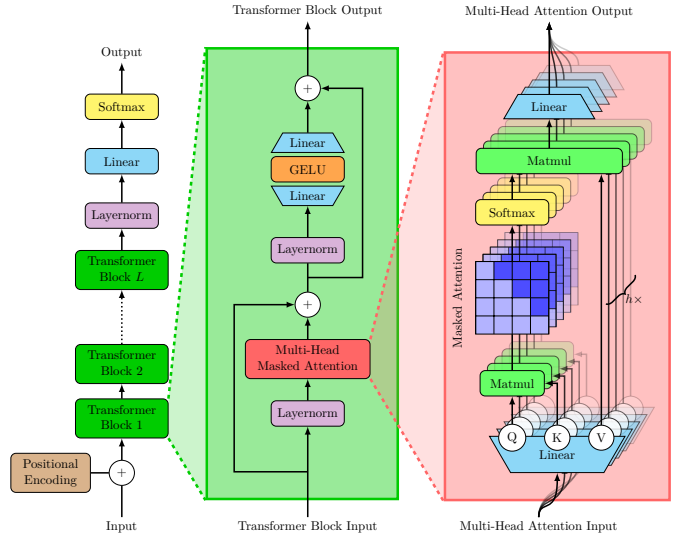
\begin{figure}[tbh]
  \centering
  \resizebox{\columnwidth}{!}{
\colorlet{trans_color}{black!20!green}
\colorlet{att_color}{red!60}
\begin{tikzpicture}[
        node distance=1.5cm and 2.5cm,
        box/.style={rectangle, rounded corners, minimum width=2cm, minimum height=0.7cm, text centered, draw},
        layernorm/.style={box, fill=violet!30},
        posenc/.style = {box, fill=brown!60 },
        softmax/.style = {box, fill=yellow!70},
        transformer/.style = {box, fill=trans_color},
        matmul/.style = {box, fill=green!60},
        gelu/.style = {box, fill=orange!70},
        attention/.style = {box, fill=att_color},
        linear/.style = {fill=cyan!40},
        linearBox/.style = {linear, box},
        linearTrap/.style = {linear, trapezium, minimum width=2cm, text centered, draw},
        linearTrapInv/.style = {linearTrap, trapezium angle=-60},
        plus/.style = {circle, draw, fill=white},
        line/.style = {very thick},
        arrow/.style = {line, -latex},
        mega thick/.style = {line width=2pt},
        zoom line/.style = {dashed, mega thick},
        zoom box/.style = {rectangle, mega thick, inner sep=0.75cm, fill opacity=0.4},
        zoom box intra/.style = {fill opacity=0.2},
    ]


    \node (input) {Input};

    \node (plus) [plus, above of=input] {+};
    \node (posenc) [posenc, left=0.5cm of plus] {\begin{tabular}{c}Positional \\Encoding \end{tabular}};

    \node (trans1) [transformer, above of=plus] {\begin{tabular}{c}Transformer \\ Block 1\end{tabular}};
    \node (trans2) [transformer, above of=trans1] {\begin{tabular}{c}Transformer \\ Block 2\end{tabular}};
    \node (transN) [transformer, above=of trans2] {\begin{tabular}{c}Transformer \\ Block $L$\end{tabular}};

    \node (layernorm) [layernorm, above of=transN] {Layernorm};
    \node (linear) [linearBox, above of=layernorm] {Linear};
    \node (softmax) [softmax, above of=linear] {Softmax};

    \node (output) [above of=softmax] {Output};

    \draw[arrow] (trans1) -- (trans2);
    \draw[arrow, dotted] (trans2) -- (transN);
    \draw[arrow] (transN) -- (layernorm);
    \draw[arrow] (layernorm) -- (linear);
    \draw[arrow] (linear) -- (softmax);
    \draw[arrow] (softmax) -- (output);

    \draw[line] (input) -- (plus);
    \draw[line] (posenc) -- (plus);
    \draw[arrow] (plus) -- (trans1);


    \node (trans_input) [right=of input] { Transformer Block Input };
    \node (trans_layernorm1) [layernorm, above=2.5cm of trans_input] { Layernorm };
    \node (trans_attention) [attention, above of=trans_layernorm1] {\begin{tabular}{c} Multi-Head \\ Masked Attention \end{tabular}};
    \node (trans_plus) [plus, above of=trans_attention] {+};
    \node (trans_layernorm2) [layernorm, above of=trans_plus] { Layernorm };
    \node (trans_linear1) [linearTrapInv, above of=trans_layernorm2] { Linear };
    \node (trans_gelu) [gelu, above=0.1cm of trans_linear1] { GELU };
    \node (trans_linear2) [linearTrap, above=0.1cm of trans_gelu] { Linear };
    \node (trans_plus2) [plus, above of=trans_linear2] {+};
    \node (trans_output) [above=1.5cm of trans_plus2] {Transformer Block Output};

    \draw[arrow] (trans_input) -- coordinate (trans_mid1) (trans_layernorm1);
    \draw[arrow] (trans_layernorm1) -- (trans_attention);
    \draw[arrow] (trans_attention) -- (trans_plus) ;
    \draw[line] (trans_plus) -- coordinate (trans_mid2) (trans_layernorm2) ;
    \draw[arrow] (trans_layernorm2) -- (trans_linear1) ;
    \draw[arrow] (trans_linear2) -- (trans_plus2) ;
    \draw[arrow] (trans_plus2) -- (trans_output) ;

    \node (trans_dot1) [coordinate, left=0.3cm of trans_attention] {};
    \draw[arrow] (trans_mid1) -| (trans_dot1) |- (trans_plus);

    \node (trans_dot2) [coordinate, right of=trans_linear2] {};
    \draw[arrow] (trans_mid2) -| (trans_dot2) |- (trans_plus2);

    \begin{scope}[on background layer]
        \node (trans_all) [zoom box, fit=(trans_mid1) (trans_plus2) (trans_linear2) (trans_dot1) (trans_attention), fill=trans_color, draw=trans_color] {};
    \end{scope}
    \draw[zoom line, draw=trans_color] (trans1.north east) -- (trans_all.north west);
    \draw[zoom line, draw=trans_color] (trans1.south east) -- (trans_all.south west);
    \fill[trans_color, zoom box intra] (trans1.north east) -- (trans1.south east) -- (trans_all.south west) -- (trans_all.north west) -- cycle;


    \node (att_input) [right=2cm of trans_input] { Multi-Head Attention Input };
    \node (att_input_coord) [coordinate, above of=att_input] {};

    \foreach \i in {4,3,2,1,0} {
        \begin{scope}[opacity={1-0.2*\i},node distance=1.5cm]
            \node (att_linear1_\i) [linearTrapInv, minimum height=1cm, above right=0.2*\i cm and 0.2*\i cm of att_input_coord] {\ifnum\i=0 Linear\else\phantom{Linear}\fi};

            \node (k_\i) [plus, above=-0.4cm of att_linear1_\i] {\ifnum\i=0 K\else\phantom{K}\fi};
            \node (q_\i) [plus, left=0.5cm of k_\i] {\ifnum\i=0 Q\else\phantom{Q}\fi};
            \node (v_\i) [plus, right=0.5cm of k_\i] {\ifnum\i=0 V\else\phantom{V}\fi};

            \node (att_matmul_K_\i) [matmul, minimum width=5em, above of=q_\i] {\ifnum\i=0 Matmul\else\phantom{Matmul}\fi};

            \begin{scope}[shift={(att_matmul_K_\i.north)}]
                \fill[blue!30] (-1,1) rectangle (1,3);
                \fill[blue!70] (-0.5,3) -- (-0.5,2.5) -- (0,2.5) -- (0,2) -- (0.5, 2) -- (0.5,1.5) -- (1, 1.5) -- (1, 3) -- cycle;
                \draw[step=0.5] (-1,1) grid (1,3);
                \draw (-1,1) rectangle (1,3); 

                \coordinate (att_grid_mid_\i) at (0,2);
                \coordinate (att_grid_south_\i) at (0,1);
                \coordinate (att_grid_north_\i) at (0,3);
            \end{scope}

            \node (att_softmax_\i) [softmax, above=2cm of att_grid_mid_\i] {\ifnum\i=0 Softmax\fi};

            \node (att_matmul_V_\i) [matmul,
                minimum width=4cm,
                above right=0.7cm and -2cm of att_softmax_\i] {\ifnum\i=0 Matmul\fi};

            \draw[arrow] (att_input.north) to[out=90, in=-90, in distance=1.2cm] (att_linear1_\i.south);

            \draw[arrow] (att_softmax_\i) -- (att_softmax_\i.north |- att_matmul_V_\i.south);

            \draw[arrow] (q_\i) -- (att_matmul_K_\i);
            \draw[arrow] (k_\i) |- (att_matmul_K_\i);
            \draw[arrow] (v_\i) -- coordinate (att_line_v_\i) (v_\i.north |- att_matmul_V_\i.south);

            \draw[line] (att_matmul_K_\i) -- (att_grid_south_\i);
            \draw[arrow] (att_grid_north_\i) -- (att_softmax_\i);

            \node (att_linear2_\i) [linearTrap, minimum width=2.5cm, above of=att_matmul_V_\i] {\ifnum\i=0 Linear\else\phantom{Linear}\fi};
            \draw[arrow] (att_matmul_V_\i) -- (att_linear2_\i);
        \end{scope}
    }

    \draw [
    decorate,
    very thick,
    decoration = {brace,
        raise=5pt,
        mirror,
        amplitude=5pt,
        }] (att_line_v_0) --  (att_line_v_4)
    node[pos=0,right=15pt,black]{$h\times$};

    \node [left of=att_grid_mid_0, rotate=90] {Masked Attention};

    \node (att_output) [above=2cm of att_linear2_0] { Multi-Head Attention Output };

    \foreach \i in {4,3,2,1,0} {
        \begin{scope}[opacity={1-0.2*\i},node distance=1.5cm]
            \draw[arrow] (att_linear2_\i) to[out=90, in=-90, in distance=0.9cm] (att_output);
        \end{scope}
    }

    \begin{scope}[on background layer]
        \node (att_all) [zoom box, inner xsep=1.5cm, fit=(att_linear1_0) (att_linear2_2), fill=att_color, draw=att_color] {};
    \end{scope}
    \draw[zoom line, draw=att_color] (trans_attention.north east) -- (att_all.north west);
    \draw[zoom line, draw=att_color] (trans_attention.south east) -- (att_all.south west);
    \fill[att_color, zoom box intra] (trans_attention.north east) -- (trans_attention.south east) -- (att_all.south west) -- (att_all.north west) -- cycle;

\end{tikzpicture}}
  \caption{Modified architecture of the GPT-2 model used in this work.}
  \label{fig:GPT2-arch}
\end{figure}

The GPT-2 model used in this work is based on Andrej Karpathy's
open-source repository \texttt{llm.c} published on GitHub in
2024~\cite{Karpathy2024-llm.c}.  The architecture of this model
largely follows that of the aforementioned original
works~\cite{OpenAI2019-GPT2-pytorch} with some minor modifications.
Figure~\ref{fig:GPT2-arch} shows the actual architecture used in this
study, composed of about a dozen of kernels (encoder, LayerNorm, GELU,
matmul \ldots), with the following key differences:
\begin{itemize}
\item Dropout layers are removed.
\item The second residual forward links to the output of the
  multi-head attention layer, instead of the one associated with the
  normalization layer.
\item The model's last layer is the Softmax function.
\item The linear layers inside the multi-head attention mechanism
  are fused with matmul functions.
\end{itemize}

We employ the largest version of GPT-2 consisting of 1558$\,$M parameters
that derive from 48 connected layers, 25 attention heads, and 1600
hidden dimensional states.  The vocabulary consists of 50,257 tokens
and the maximum context length is 1024 tokens.  We reutilize OpenAI's
GPT-2 weights (32-bit precision) publicly available at its Hugging
Face repository~\cite{OpenAI2019-GPT2-pytorch-model} and continue the
pre-training of GPT-2 over the TinyShakespeare
dataset~\cite{Karpathy2024-TinyShakespeare} composed of 305,260
tokens.  The AdamW optimization function employs the following
parameters: $\alpha = 10^{-4}$, $\beta_1 = 0.9$, $\beta_2 = 0.999$,
$\epsilon = 10^{-8}$, and $\lambda = 0$.  Finally, we reduce the context length and batch size to 256 and 4 tokens, respectively, to ensure that the model fits within the 64GB of HBM2 memory available on a single NVIDIA H100 GPU. \ab{With this configuration, the theoretical memory required for the model
is $\approx 45\,\text{Gb}$, which fits in the GPU memory with margin to
accommodate any temporary buffers needed in some kernels.}

\subsubsection{GPT-2 fork-join parallelisation}
\begin{listing}[htb]
\small  
\begin{minted}[frame=single, linenos, numbersep=5pt]{C}
const int o_start = slh.dims_starts[0];
const int o_end = slh.dims_stops[0];
if (dbias != NULL) {
  #pragma omp parallel for
  for (int j = o_start; j < o_end; j++) {
    for (int b = 0; b < B; b++) {
      const float* dout_b = dout + b*T*OC;
      float wrk = 0.0f;
      #pragma omp simd simdlen(SIMDLEN) \
        reduction(+: wrk)
      for (int t = 0; t < T; t++)
        wrk += dout_b[j + t*OC];
      dbias[j] += wrk;
    }
  }
}
#pragma omp parallel for
for (int j = o_start; j < o_end; j++) {
  for (int k = 0; k < B*T; k++) {
    const float wrk = dout[j + k*OC];
    for (int i = 0; i < C; i++)
      dweight[i + j*C] += inp[i + k*C]*wrk;
  }
}
\end{minted}
\caption{Matmul kernel over bias and weights gradients during the
  backward pass parallelized with OpenMP adopting a fork-join
  strategy.}\label{code:matmul-fork-join}
\end{listing}
It is worth noting that Karparthy's original version of GPT-2 version
serves an educational purpose and, therefore, is maintained simple and
readable; consequently, its computational kernels may not offer the
maximum performance.  The source code in \cite{Karpathy2024-llm.c} is
written in C and only comes with a handful of kernels parallelized
with OpenMP \texttt{parallel for} loops: matmul, attention, and
Softmax functions.

We have extended the \texttt{parallel for} construct to every kernel
and complemented it with the \texttt{simd} directives where necessary
as shown in Code~\ref{code:matmul-fork-join} corresponding to the
matmul function.  In the majority of kernels, our fork-join
parallelization is carried out along both the batch and context
dimensions by using the \texttt{collapse} clause over these two loops.
The attention kernels can be further parallelized along both the
attention head and head size dimensions.

The matmul kernels found in the original source code become the
computational bottleneck of the pre-training phase of GPT-2.  For this
reason, we have firstly separated the computation associated with the
bias (see lines~3--16 of Code~\ref{code:matmul-fork-join}) and,
secondly, we have rewritten the remaining part (lines~17--24) in terms
of a general matrix-matrix multiplication (GEMM).  This refactoring
not only simplifies and improves the readability of the original
source code, but it also allows the compiler to apply further
optimizations.  Moreover, one can now replace these in-house kernels
by calls to highly tuned GEMM functions from third-party BLAS
libraries.  \pmf{While this speeds up multi-head masked attention operations significantly, as well as other linear transformations both inside and outside the Transformer block, it does not bring the same degree of performance as the popular FlashAttention algorithm~\cite{Dao2022-FlashAttention}.}

\subsubsection{GPT-2 task parallelization}
\begin{listing}[htb]
\small
\begin{minted}[frame=single, linenos, numbersep=5pt]{cpp}
const int Vp_gran = Vp/OC_SPLIT;
#pragma oss taskloop grainsize(1) \
  in({grads_acts_logits[b*T*Vp + t*Vp], \
     b=0;B:B_GRAN, t=0;T:T_GRAN}) \
  in({acts_lnf[b*T*C + t*C], \
     b=0;B:B_GRAN, t=0;T:T_GRAN}) \
  out(grads_wte[o*C; Vp_gran*C])
for (int o = 0; o < Vp; o += Vp_gran) {
  struct SliceHandler slh;
  slh.dims_starts[0] = o;
  slh.dims_stops[0] = MIN(o + Vp_gran, Vp);
  matmul_params_backward(grads_wte,
                         nullptr,
                         grads_acts_logits,
                         acts_lnf,
                         slh);
}
\end{minted}
\caption{Call to the function containing the matmul kernel of
  Code~\ref{code:matmul-fork-join} parallelized with OmpSs-2
  tasks.}\label{code:matmul-task}
\end{listing}
The declaration of both starting and ending task indexes along the
output channel is shown at lines~1--2 of
Code~\ref{code:matmul-fork-join}.  When using OmpSs-2 tasks, we split
the kernel along this dimension and deactivate the pragma directives
corresponding to the fork-join implementation described in the
previous section.

Code~\ref{code:matmul-task} illustrates the calls to the matmul kernel
during the backward pass.  After defining the granularity associated
with the vocabulary dimension (i.e., output channel) at line~1, we
generate the corresponding tasks with the aid of our slice handler
structure \texttt{slh}.  We require the use of multidependencies over
the batch and context dimensions to ensure the correct execution of
tasks since kernels are parallelised over these dimensions.  We opt to
fix the batch granularity to unity, that is \texttt{B\_GRAN=1} (recall
that the batch size is set to 4), and allow for different
granularities across the context length and output channels adjusting
the variables \texttt{T\_GRAN} and \texttt{OC\_SPLIT}, respectively.

Our task implementation follows a true data-flow exempt of explicit or
implicit barriers.  Nevertheless, after a certain number of training
iterations (e.g., 10), we do use \texttt{taskwait} clauses to perform
the validation of the model and execute an inference test.  Finally,
we employ the \texttt{taskiter} construct published in an earlier
work~\cite{Alvarez2023-taskiter} on top of the training iterative loop
to minimize overheads related to task creation, scheduling, and
dependency management.

\subsection{HPCCG benchmark}
The HPCCG application~\cite{Heroux2017} is a simplification of the popular HPCG benchmark~\cite{Dongarra2016} used by the TOP500 supercomputer website.  Both of them were developed within the Mantevo project~\cite{Heroux2009} and employ the conjugate gradient (CG) iterative method to numerically solve a sparse symmetric linear system of equations~\citep{Saad2003} of the form ${\bf A} \cdot {\bf x} = {\bf b}$, where ${\bf A}$ and ${\bf b}$ is the known sparse matrix and vector, respectively, and ${\bf x}$ refers to the solution vector.  ${\bf A}$ is encoded in the compressed sparse row (CSR) matrix format and, in the case of the HPCCG used in this work, no preconditioning is applied to it.  The most relevant computational kernels are: (i) the sparse--matrix vector multiplication (SpMV), (ii) the vector update (\texttt{axpy} and \texttt{scale}), and (iii) the dot product (\texttt{dot}).  Finally, HPCCG supports shared-memory parallelism via OpenMP parallel for directives.

In our previous work~\cite{Martinez2023-HLAM} we further developed HPCCG to provide task-based parallelism via OpenMP and OmpSs-2 directives.  This new data-flow strategy, although requires slightly more effort from the programmer's perspective, enables the overlapping of computation and communication phases of the algorithm thereby improving its performance.  All the extensions we made to HPCCG code base are publicly available at Code Ocean~\citep{martinez2022-HLAM} under the name HLAM (hybrid linear algebra methods).  We refer the reader to this repository and the code snippets present in our published paper~\cite{Martinez2023-HLAM} describing the two aforementioned parallel implementations on multicore CPUs.

\section{Related work}\label{sec:related}
Efficient programming of heterogeneous systems remains a challenging area that has motivated extensive research in programming models, runtime systems, and interoperability strategies. This section reviews key developments in task-aware APIs, automatic data management mechanisms, portability across platforms, and interoperability between different programming models.  Our work builds upon these efforts by addressing the orchestration and combination of multiple programming models and APIs within unified task-based applications.


\subsection{Automatic data management and directory/cache mechanisms}
Several task-based runtimes have introduced directory/cache (D/C) mechanisms to automate data movement between hosts and accelerators, aiming to improve programmability and performance. StarPU~\cite{Augonnet2011-StarPU} and OmpSs-2~\cite{Duran2011-OmpSs} are prominent examples, providing unified runtimes that manage both data transfers and task scheduling in heterogeneous systems. KAAPI~\cite{Gautier2007-KAAPI} and IRIS~\cite{Kim2021-IRIS} offer similar solutions. These mechanisms allow runtimes to transparently handle data placement and movement based on task annotations. However, their transparency may limit their interoperability with external APIs, and can lock down users to the APIs natively supported by the runtime's backend.

\subsection{Portable programming models}
There is a substantial body of research on the trade-offs between portability and performance in programming models such as OpenCL~\cite{OpenCL}, SYCL~\cite{SYCL}, OpenACC~\cite{OpenACC}, and OpenMP offload. These studies compare model productivity and efficiency across various architectures, including CPUs, GPUs, and specialized accelerators. For example, Memeti et al.~\cite{Memeti2017-benchmarking} benchmarked OpenMP, OpenACC, OpenCL, and CUDA with respect to programming effort, performance, and energy use, showing that higher performance often comes at the cost of increased complexity. Reguly~\cite{Reguly2023-evaluating} examined SYCL's performance portability on bandwidth-bound applications, demonstrating that achieving optimal results still requires tuning for each specific target. Our approach is distinct in that we address how both portable and vendor-specific APIs can be combined and orchestrated effectively, rather than evaluating them in isolation.

\subsection{Interoperability across programming models}
Interoperability between different programming models is a growing topic of interest. For example, Korakitis et al.~\cite{Korakitis2022-towards} proposed an interoperability mechanism between OmpSs-2 and OpenACC, demonstrating that composability among different pragma-based programming models can enable more flexible and efficient use of heterogeneous systems. In previous research, we find Converse~\cite{Converse} was a proposed interoperability layer oriented at having multiple parallel programming models co-exist when linked into a single application, if every runtime is implemented on top of Converse semantics. Similarly, MetaChaos~\cite{MetaChaos} was a proposed layer which would allow sharing data structures among HPF and p\CC{}, which enabled a program to be written in a combination of both programming languages.
In general, we find prior work in this area has typically focused on specific pairs of programming models or required extensive adaptations of the models to be interoperated. In contrast, our methodology generalizes interoperability to any data-flow tasking model and offloading API, enabling broader and more scalable solutions for heterogeneous computing.


\section{Task-aware libraries for heterogeneous computing}\label{talibs}
Combining multiple accelerator programming models within a single application introduces significant challenges at both the software and runtime levels. Each accelerator API like CUDA, SYCL, or Triton defines its own kernel language and exposes a custom runtime interface for managing memory transfers, kernel launches, and synchronization. As a result, applications must first implement accelerator kernels separately, using the appropriate language and compilation toolchain. These kernels are then compiled independently and linked into the application as external modules.

While this modular structure enables flexibility, orchestrating the execution of kernels written in different models is far from trivial. The most straightforward approach is to treat each accelerator phase as a monolithic block, enforcing full synchronization between them: e.g., by using global barriers or device-wide synchronization calls. This guarantees correctness but prevents overlapping computation and communication, thereby underutilizing system resources.

Achieving finer-grained parallelism and improving overlap across devices requires adopting a more advanced execution model. In this context, models like OpenMP and OmpSs-2 provide a natural way to express the application as a graph of coarse-grained tasks, each representing a distinct computation or offloading operation. These tasks form a directed acyclic graph (DAG), where dependencies between them are automatically inferred and enforced by the runtime. In principle, this approach allows heterogeneous kernels (written in CUDA, SYCL, Triton \ldots) to coexist within a single unified execution framework.

However, a key limitation arises when integrating non-blocking accelerator operations within tasks. Traditional runtimes require that a task does not complete until all internal operations (e.g., memory copies or kernel executions) have finished. This means that developers must introduce explicit synchronization points (such as
\texttt{cudaStream\-Synchronize()}
or \texttt{q.wait()}) at the end of each task to avoid data races. These synchronizations block the CPU core running the task, preventing the runtime from scheduling other ready tasks and thus leading to underutilization of CPU resources.

Task-aware libraries address this problem by decoupling task completion from device synchronization. They enable accelerator operations to be issued from within tasks without requiring explicit blocking calls at the end of the task. In the following, we describe how the task-aware libraries TACUDA and TASYCL track the lifecycle of non-blocking operations and defer dependency release until all associated events have completed, allowing CPU cores to remain available and improving the overall efficiency of heterogeneous task-based applications.

To understand the specific challenges that task-awareness solves, let us take a look at a specific example from a hypothetical application using OpenMP tasks and CUDA kernels, showcased in Code~\ref{code:cuda-openmp-nota}.

\begin{listing}[tb]
\small
\begin{minted}[frame=single, linenos, numbersep=5pt]{C}
void gpu_task_0(...) {
  cudaMemcpyAsync(..., stream);
  launch_kernel<<<..., stream>>>(x, y);
  cudaMemcpyAsync(..., stream);
  cudaStreamSynchronize(&stream);
}
int main() {
  [...]
  for (int i = 0; i < N; ++i) {
    #pragma omp task depend(out: x[i]) // C0
    cpu_task_0(&x[i]);
    #pragma omp task depend(in: x[i]) \
      depend(out: y[i]) // G0
    gpu_task_0(&x[i], &y[i]);
    #pragma omp task depend(in: x[i]) // C1
    cpu_task_1(&x[i]);
    #pragma omp task depend(in: y[i]) // C2
    cpu_task_2(&y[i]);
  }
  [...]
}
\end{minted}
\caption{Sample OpenMP+CUDA application.}\label{code:cuda-openmp-nota}
\end{listing}

\begin{figure}
    \centering
    \subfloat[Without TACUDA.]{\includegraphics[width=\columnwidth]{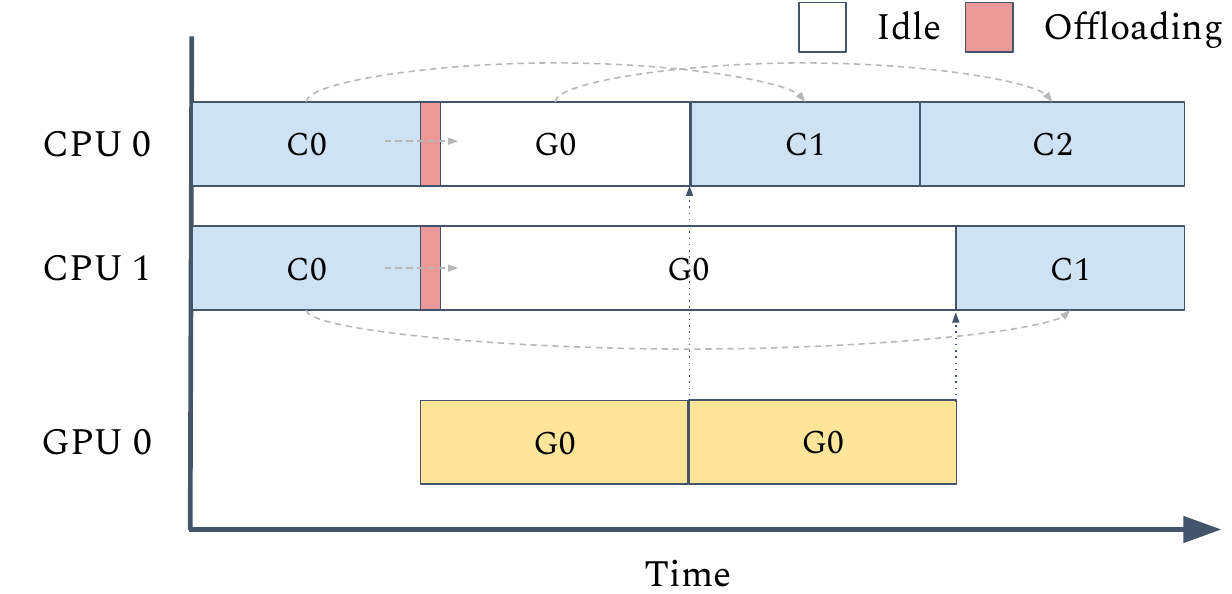}}
    \\
    \subfloat[With TACUDA.]{\includegraphics[width=\columnwidth]{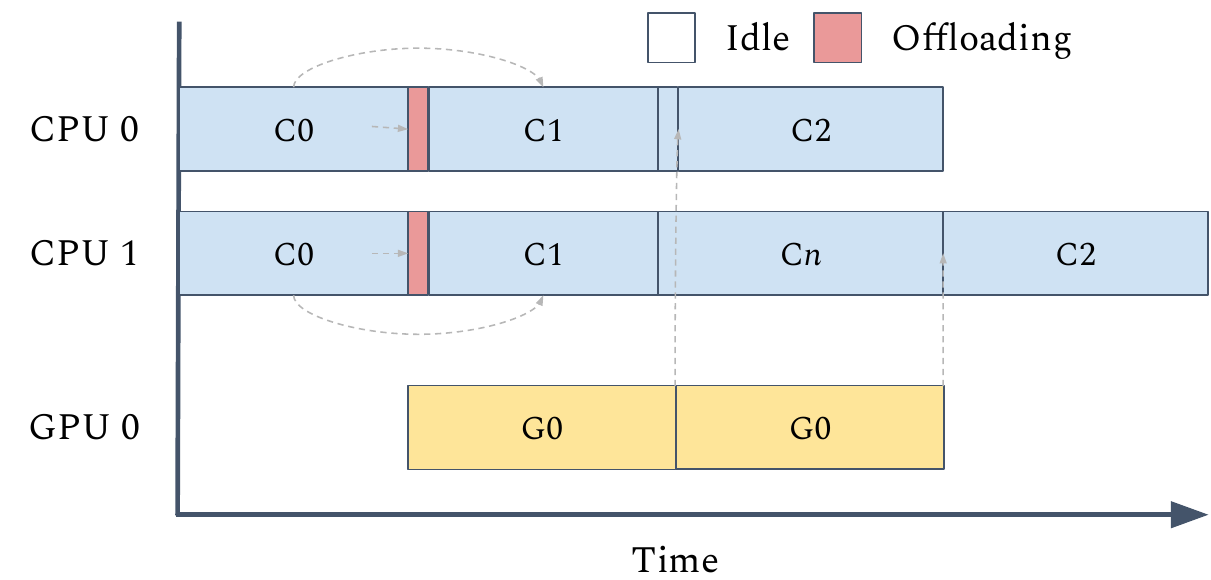}}
    \caption{Application trace of Code~\ref{code:cuda-openmp-nota} (a) without and (b) with task-awareness. Dashed lines represent task dependencies, while dotted lines represent device-host synchronization.}
    \label{fig:trace-cuda-openmp}
\end{figure}

In this application, we aim at offloading some operations to our CUDA GPU, and thus we
wrap them in tasks. Successive operations, which depend on the results of
the offloaded kernels, are synchronized through data dependencies, ensuring that they will not
be executed until the results are available back in the host. However, to guarantee correctness,
we cannot let the offloading tasks end before the accelerator has finished, hence
the use of \texttt{cudaStreamSynchronize()} function. This application could result in a trace
similar to the one shown in Fig.~\ref{fig:trace-cuda-openmp}~(a), where all CPUs end up blocked in CUDA calls waiting for the accelerator to finish, even if there are other eligible tasks that could be running in the CPU during the same time period. To summarize, we are wasting CPU resources waiting for device synchronizations only to guarantee that task data dependencies are respected. 

However, if instead of plain CUDA we use the TACUDA library, we would obtain a trace similar to the one in Fig.~\ref{fig:trace-cuda-openmp}~(b), in which we do not waste CPU resources; yet, we maintain the correctness of the data-flow program by ensuring that data dependencies are not released until the accelerator has finished the enqueued computations.

\begin{figure}[hbt]
    \centering
    \includegraphics[width=\columnwidth]{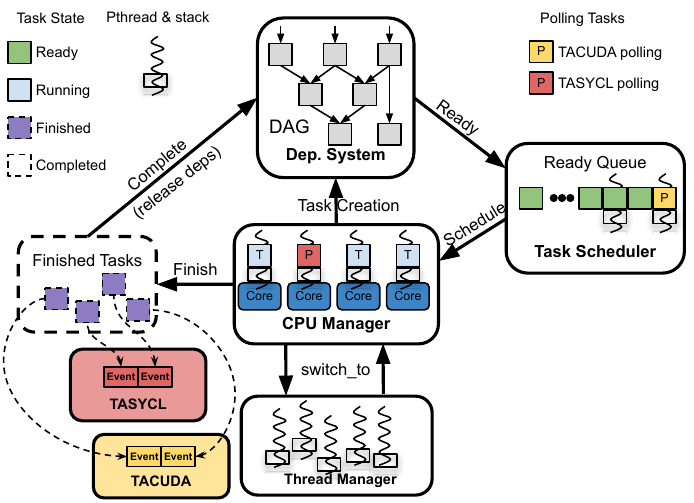}
    \caption{Integration of task-aware libraries within a task-based data-flow programming model.} 
    \label{fig:tasycl_tacuda_polling}
\end{figure}

All task-aware libraries, including TACUDA, TASYCL, TAMPI, and TAGASPI, do share a common implementation architecture designed to support both blocking and non-blocking APIs within task-based data-flow programming models. This mechanism is illustrated in Fig.~\ref{fig:tasycl_tacuda_polling}. For blocking operations, the library transparently replaces the blocking call with its non-blocking counterpart and suspends the calling task. A dedicated polling task, whose polling frequency can be configured by the user, periodically checks the status of all in-flight operations. Once an operation completes, the corresponding suspended task is resumed and reinserted into the scheduler's ready queue. For non-blocking operations, the runtime associates a counter with each task. When a task issues a non-blocking operation, the counter is incremented; upon task completion, the task's data dependencies are not released if the counter is non-zero. The polling task also monitors these non-blocking operations, and decrements the counter as operations complete. Once the counter reaches zero, the runtime safely releases the task's dependencies. This mechanism ensures correct synchronization without blocking CPU cores, enabling efficient overlap of host and accelerator work.

\ab{
Alternative mechanisms to the polling task approach exist through the
combination of OpenMP detach tasks with event callbacks (e.g., SYCL's
\texttt{host\_task} or CUDA's \texttt{cudaLaunchHostFunc}). In this
approach, a callback is registered in the device queue
that ``fulfills'' the detached task upon completion,
thereby eliminating the need for active
polling. While this callback-based mechanism can reduce the overhead associated
with polling tasks, it incurs additional implementation complexity through the
management of OpenMP detach handles. Furthermore, callback functions execute
within runtime threads managed by the device driver (CUDA, SYCL, etc.), which
operate independently of the OpenMP runtime thread pool, potentially introducing
contention. The behavior of these driver-managed threads constitutes an
implementation detail that varies across platforms. For instance, CUDA may
serialize callbacks from independent streams even in the absence of explicit
ordering constraints \cite{cudatoolkit_launchhostfunc}, thereby introducing
delays in dependency release. In contrast, the polling task mechanism maintains
tight integration with the tasking runtime, offering more predictable behavior
across different implementations and enabling performance tuning through
polling frequency adjustments.
}


\subsection{Triton}

Unlike the other mechanisms explored in this work, Triton is not C/\CC{}-based.  Its kernels are written in a Python-based DSL, must reside in separate Python files, and cannot be directly invoked from native \CC{} applications. To integrate them with such applications, we perform ahead-of-time (AoT) compilation targeting the desired backend platform.  It is worth mentioning that Triton itself only generates the (CUDA) kernels and therefore does not provide APIs to manage their execution or synchronization. Instead, these responsibilities are handled via the native backend API like CUDA, which is ultimately responsible for launching and synchronizing the kernels. As a result, there is no need for a dedicated task-aware Triton library and we simply leverage the corresponding task-aware library that matches the backend API used by Triton (TACUDA when targeting NVIDIA GPUs) to ensure seamless integration with the task-based runtime.  More specifically, when compiling for CUDA GPUs, each kernel outputs a CUBIN blob. These blobs are loaded during application startup via the CUDA Driver API using \texttt{cuModuleLoadData}.  Next, \CC{} applications can invoke such kernels using \texttt{cuLaunchKernel}. Finally, we pass a TACUDA-managed stream to \texttt{cuLaunchKernel} to make them task-aware.

For Triton, all GPU kernels used in the GPT-2 pre-training phase and the HPCCG benchmark have been fully rewritten in the Triton programming model to take advantage of its fine-grained, tile-based execution strategy. Unlike directive-based approaches, Triton requires explicit kernel definitions in Python using its specialized DSL, which provides abstractions for manipulating memory tiles and optimizing data locality. Each kernel is structured to operate over blocks of data that map naturally to the GPU's hardware hierarchy, enabling efficient use of shared memory and minimizing memory bandwidth pressure. In the case of GPT-2, operations such as matrix multiplication and layer normalization are tiled along batch and context dimensions, exposing parallelism across attention heads and hidden states. Similarly, for HPCCG, sparse matrix-vector multiplication is restructured to process contiguous rows of the matrix in parallel, while caching relevant vector slices in shared memory to minimize redundant global memory accesses. This explicit control over memory movement and parallelism allows Triton kernels to approach the performance of hand-optimized device code, while still benefiting from portability across backend targets.

\subsection{OpenMP offload}
We employ the \texttt{target} clause integrated in the OpenMP programming model to offload kernels to the GPU.  By default, GPU kernels block the corresponding CPU thread calling it.  To prevent this from happening, we integrate blocking OpenMP GPU kernels within OmpSs-2 (or OpenMP) tasks.  It is also necessary to substitute the pragma directives previously mentioned in Section~\ref{sec:background}.  For example, line~17 of Code~\ref{code:matmul-fork-join} is equivalent to:
{\small
\begin{minted}[]{C}
#pragma omp target teams distribute parallel for
\end{minted}
}
\noindent which allows us to make a fair comparison between different
OpenMP kernel transformations on CPUs and GPUs.  However, the generated
GPU code offers subpar performance.  To partially overcome this
limitation, and thus increase the computational speed, we collapse
the loops along \texttt{j} and \texttt{k} dimensions (lines~18--19 of
Code~\ref{code:matmul-fork-join}),
that is, we parallelize the GEMM operation over the output channel
(e.g., vocabulary or hidden dimensional state), batch, and context
dimensions, which requires atomic updates over variable
\texttt{dweight} at line~22.

\begin{listing}[htb]
\small
\begin{minted}[frame=single, linenos, numbersep=5pt]{C}
#pragma omp target teams distribute parallel for
for (int j = 0; j < rowBsA; ++j) {
  const int rIdx0 = dA_rowIdx[rowIdx+j];
  const int nnz = dA_rowIdx[rowIdx+j+1] - rIdx0;
  double wrk = 0.0;
  for (int k = 0; k < nnz; ++k) {
    wrk += dA_values[rIdx0+k]*
           d_x[dA_columns[rIdx0+k]];
  }
  d_Ax[j+rowIdx] = wrk;
}
\end{minted}
\caption{Sparse matrix-vector multiplication (SpMV) based on CSR format and parallelized via OpenMP target directives.}\label{code:spmv}
\end{listing}
With regard to the HPCCG benchmark, the OpenMP target implementation for GPUs is straightforward for the three type of kernels found in this application as shown in Code~\ref{code:spmv} corresponding to the sparse matrix-vector multiplication.  Note that this code excerpt works on horizontal tiles of the sparse matrix \texttt{A} and consecutive regions of the output vector \texttt{d\_Ax} by using the starting row index (\texttt{rowIdx}) and row block sizes (\texttt{rowBsA}) variables hence allowing us to combine both target and task directives within the same code.

\subsection{SYCL}
SYCL kernels are submitted to a device queue using the \texttt{parallel\_for} construct, and operate over multi-dimensional ranges to expose parallelism. As shown in Code~\ref{code:spmv_sycl}, the sparse matrix-vector multiplication (SpMV) kernel used in HPCCG is offloaded using a 1D range, where each work-item processes a separate row of the matrix. The kernel traverses the compressed sparse row (CSR) data structures to compute the dot product of each row with the input vector, storing the result in the corresponding entry of the output vector.
\begin{listing}[htb]
\small
\begin{minted}[frame=single, linenos, numbersep=5pt]{cpp}
sycl::event ev = Q.parallel_for<class spmvA>(
    sycl::range<1>(rowBsA), [=](sycl::id<1> j) {
  const int rIdx0 = dA_rowIdx[rowIdx + j];
  const int nnz = dA_rowIdx[rowIdx + j + 1] - rIdx0;
  double wrk = 0.0;
  for(int k = 0; k < nnz; ++k) {
    wrk += dA_values[rIdx0 + k] *
           d_x[dA_columns[rIdx0 + k]];
  }
  d_Ap[j + rowIdx] = wrk;
});
\end{minted}
\caption{Sparse matrix-vector multiplication (SpMV) based on CSR format and parallelized via SYCL.}%
\label{code:spmv_sycl}
\end{listing}

All GPU kernels from the GPT-2 pre-training phase and the HPCCG benchmark have been ported from their CUDA versions. This translation is generally straightforward, as the SYCL programming model exposes a similar hierarchical parallelism abstraction through ND-range kernels, which closely mirrors CUDA's grid-block-thread execution structure. Kernel logic and memory access patterns remain largely unchanged, allowing for direct reuse of indexing strategies and loop structures.

All other kernels required for GPT-2, including linear layers, softmax, and normalization operations, follow a similar pattern: they are re-expressed as ND-range SYCL kernels and offloaded through device queues. Since SYCL kernels are inherently non-blocking, they integrate naturally with the TASYCL library, which ensures correct synchronization and dependency tracking through the task-aware mechanism described in Section~\ref{talibs}.

\ab{
Code~\ref{code:tasycl} illustrates the basic usage of the TASYCL library for
integrating SYCL kernels with task dependencies. A SYCL queue is obtained
from the pool via \texttt{tasyclGetQueue} and later returned with
\texttt{tasyclReturnQueue}. The function \texttt{tasyclSynchronizeEventAsync}
replaces the \texttt{ev.wait()} blocking call, and instead
binds the task completion with the SYCL event \texttt{ev},
ensuring that dependent tasks are not released until the kernel execution
finishes.
}

\begin{listing}[htb]
\small
\begin{minted}[frame=single, linenos, numbersep=5pt]{cpp}
#pragma omp task depend(in: x[0:n], inout: y[0:n])
void axpy(int n, double a, double *x, double *y) {
  sycl::queue q = tasyclGetQueue();
  sycl::event ev = q.parallel_for<class axpy>(
    sycl::range<1>(n), [=](sycl::id<1> i) {
      y[i] += a * x[i];
    });
  tasyclSynchronizeEventAsync(ev);
  tasyclReturnQueue(q);
}
\end{minted}
\caption{axpy kernel using TASYCL within an OpenMP task.}%
\label{code:tasycl}
\end{listing}

\ab{
\subsection{PoCL nOS-V device}

Portable OpenCL (PoCL) \cite{jaaskelainen_pocl_2015} is an Open Source
implementation of the OpenCL standard.
Like other OpenCL implementations, it can be used as a SYCL backend through Intel's Unified Runtime.
It currently supports various CPU
architectures as well as CUDA (through libCUDA) and Intel GPUs (through Level
Zero). The CPU driver is implemented as work-stealing threads built using
pthreads. This kind of runtime maps well to the nOS-V model, and the modularity
of PoCL allows the creation of a new device based on that model. Our nOS-V PoCL
device is built on top of the pthread one, replacing the threads with nOS-V
tasks and changing the synchronization primitives to use nOS-V. This methodology
is analogous to what was done in
\texttt{libompv}~\cite{navarro_synergizing_2025}.
}


\section{Experimental evaluation}
\label{sec:evaluation}
This section evaluates the performance and interoperability benefits of our approach using two representative applications, HPCCG and GPT-2, on heterogeneous platforms. We begin by describing the hardware systems used for our experiments, including both general-purpose and GPU-accelerated nodes (Section~\ref{sec:eval:hardware}). We then present the evaluation results in two steps. First, we focus on GPU-based execution (Section~\ref{sec:eval:gpu}) and compare monolithic kernels with data-flow executions. We evaluate four variants: OpenMP offload, SYCL, Triton, and a hybrid \ab{``mixed''} version that combines kernels written in \ab{all three} programming models \ab{(\ref{sec:appendix:mixed} contains detailed tables showing what model was used for each kernel)}. We further compare implementations using custom GPU kernels with versions leveraging vendor-optimized libraries for performance-critical operations (SpMV in HPCCG and \pmf{GEMM} in GPT-2). Next, we analyze multi-core CPU executions (Section~\ref{sec:eval:multicore}), comparing fork-join and task-based models, and evaluate the impact of custom and vendor libraries for said kernels. Finally, we study task granularity and oversubscription effects between OpenMP, OmpSs-2, and SYCL.

\subsection{HPC systems and code setup}
\label{sec:eval:hardware}
Our evaluations are conducted on two different hardware
architectures available at the Barcelona Supercomputing Center (BSC).
The first one is the general-purpose MareNostrum 5 supercomputer
partition, namely MN5-GPP, whose compute nodes are composed of two
Intel Xeon Platinum 8480+ CPUs with a total of 112~cores running at
2~GHz.  Each CPU integrates 105~MiB of L3 cache, and supports both the
AVX-512 and AMX families of hardware accelerated instructions.  The
system memory is 256~GB based on DDR5 with a theoretical bandwidth of
307.2~GB/s per socket.  The second computational infrastructure entitled MN5-ACC integrates two Intel Xeon Platinum 8460Y+ CPUs per compute node with a total of 80 cores at 2.3~GHz, same L3 cache, and twice the system memory of the general partition with the same technical characteristics.  The two CPUs are connected to four NVIDIA Hopper H100 GPUs equipped with 64~GB of HBM2 achieving 2.02~TB/s. 

For the sake of simplicity, all experiments are carried out on a single compute node. We adjust the problem sizes of both benchmarks so that their working sets fit entirely within the host and device memory, allowing us to focus on the behavior and interoperability of heterogeneous accelerator APIs without the added complexity of distributed execution.
\ab{We also use the default thread affinity settings to keep the setup straightforward.}
While our setup avoids communication between nodes, task-aware communication libraries such as TAMPI can be seamlessly combined to support distributed scenarios.

\ab{
\subsection{Methodology}
\label{sec:eval:methodology}

For our evaluations, we measured individual iteration times, discarding the
initial and final parts of the programs which were devoted to memory transfers.
In HPCCG, every execution performed 5 repetitions of the iterative algorithm, each doing
50 iterations. The first repetition and the first 10 iterations were discarded
as ``warmup''. The same approach was used GPT-2, were 10 pre-training steps were
done, discarding the first one. Each execution was repeated 5 times on different
compute nodes.


For the GPU versions, NVTX was used to annotate the iterations, which could be
exported using NVIDIA's nsight systems, along with kernel execution. This data
was then processed to find the top kernels contributing to the execution time.
%
}

\subsection{GPU evaluation}
\label{sec:eval:gpu}
We evaluate the GPU execution using the GPT-2 pre-training and HPCCG applications across four kernel programming models: OpenMP offload, SYCL, Triton, and a hybrid ``mixed'' version (combining kernels from the three models). Each application is executed using both a monolithic fork-join implementation and a task-based version composed of coarse-grained GPU tasks (4 tasks for both GPT-2 and HPCCG). For each configuration, we analyze the top kernels contributing to execution time and evaluate the effect of using custom versus vendor-optimized kernels for performance-critical operations (matrix multiplication and SpMV).

\begin{figure}[htpb]
    \centering
    \includegraphics[width=\columnwidth]{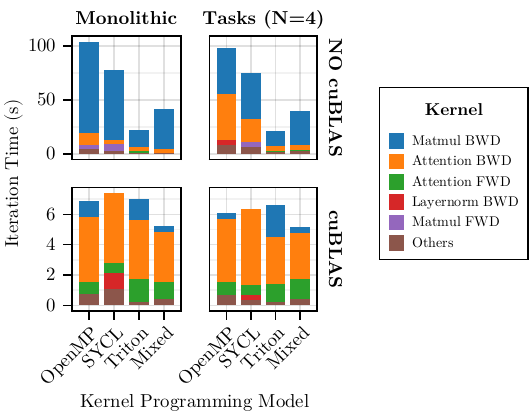}
    \caption{GPU evaluation of GPT-2.
We compare monolithic and task-based executions ($N=4$ tasks) for four kernel programming models (OpenMP offload, SYCL, Triton, and a combination of all the others ``Mixed'')
both with and without cuBLAS. Each bar reports the top three kernels by duration for a single iteration of the model (context length of 256 tokens and 1.5\,B parameters) executed on a single H100 GPU.  Bars are stacked with the time contribution of each kernel.
\ab{Each iteration processed a total of 1024 tokens (256 tokens in context $\times$ 4 token batches), which corresponds
to $203\,\text{tokens}/s$ in our best result (Tasks--Mixed--cuBLAS).}
}%

    \label{fig:kernels_gpt2_gpu}
\end{figure}

Figure~\ref{fig:kernels_gpt2_gpu} shows results for GPT-2 pretraining. In this workload, the task-based and fork-join versions exhibit comparable performance across all kernel models, indicating that the overheads of GPU task orchestration are negligible at this granularity. Across all configurations, the dominant contributors to iteration time are the matrix multiplication and attention kernels. When vendor-optimized BLAS libraries are used for the matmul kernel (cuBLAS), performance improves significantly in all cases, with the ``mixed'' version achieving the best overall performance. This hybrid approach leverages model-specific strengths, e.g. Triton for attention kernels and SYCL or OpenMP for other kernels, demonstrating the benefit of composing kernels written in different programming models within a task-based framework.

\begin{figure}[htpb]
    \centering
    \includegraphics[width=\columnwidth]{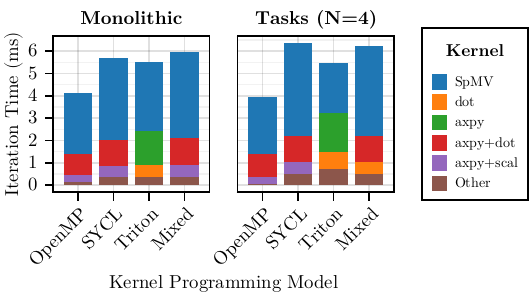}
    \caption{
    GPU breakdown for the HPCCG application with $256^{3}$ points executed on a single H100 GPU. For each kernel we report the three most time-consuming kernels, stacked to show their contribution to a full iteration.
}%

    \label{fig:kernels_hpccg_gpu}
\end{figure}

The HPCCG results shown in Fig.~\ref{fig:kernels_hpccg_gpu} align with the trend already observed for GPT-2: the monolithic and task-based executions achieve virtually identical iteration times, with the task version marginally outperforming the monolithic one for the Triton and ``Mixed'' configurations. The performance is dominated by the memory-bound SpMV kernel, whose low arithmetic intensity and irregular access pattern leave little headroom for further optimization. For this reason we omit the cuBLAS variants: replacing our custom kernels with the vendor implementation yields iteration times within measurement noise. These results confirm that, when task granularity is chosen appropriately, a data-flow execution can match and, in some cases slightly surpass, a monolithic version.

\subsection{Multi-core evaluation}
\label{sec:eval:multicore}

This section evaluates the performance implications of combining multiple runtime systems in CPU-only environments, focusing on both fork-join and task-based parallel executions. Our goal is to quantify the impact of runtime interference and evaluate how a unified tasking substrate, such as nOS-V, can mitigate oversubscription and enable efficient resource sharing across runtimes.

\begin{figure}[htpb]
    \centering
    \includegraphics[width=\columnwidth]{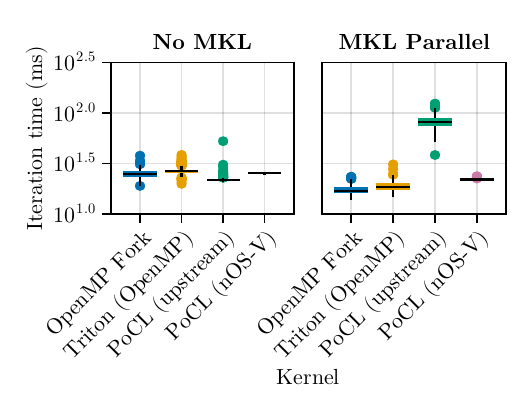}
    \caption{
        CPU evaluation of HPCCG. Mixing fork-join models on HPCCG ($256^3$ points) with MKL (note that MKL is only used for the SpMV kernels).
    }
    \label{fig:kernels_seq_hpccg_cpu}
\end{figure}

\paragraph{HPCCG fork-join execution}
Figure~\ref{fig:kernels_seq_hpccg_cpu} shows the iteration time for the HPCCG benchmark executed using different kernel and runtime configurations. In the left panel, we run the application using OpenMP fork-join, sequential Triton kernels launched from OpenMP, and SYCL kernels executed via PoCL (both the upstream and the nOS-V integrated version). OpenMP and Triton rely on LLVM's \texttt{libomp}, while PoCL maintains its own internal thread pool. All configurations perform similarly, with PoCL~(nOS-V) showing the lowest variability.

The right panel introduces Intel's MKL library for the sparse matrix-vector multiplication (SpMV) kernel, which internally uses \texttt{libomp}. When combined with OpenMP or Triton, MKL executes efficiently, as all components share the same threading runtime. However, combining MKL with upstream PoCL introduces significant performance degradation due to runtime interference and thread oversubscription. This issue is effectively resolved when both \texttt{libomp} and PoCL are configured to use nOS-V, enabling cooperative scheduling and restoring performance to baseline levels.

\begin{figure*}[htpb]
    \centering
    \includegraphics{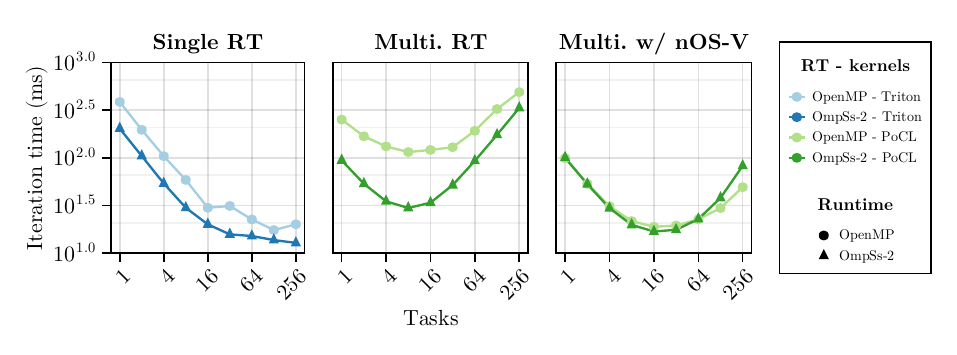}
    \caption{
        HPCCG execution on CPU.  Task-based execution with varying task granularities and runtime combinations. 
        The left panel uses sequential Triton kernels while the middle and right panels use SYCL kernels via PoCL (upstream and nOS-V, respectively).
    }
    \label{fig:kernels_hpccg_cpu_gran_multi}
\end{figure*}

\paragraph{HPCCG task-based execution}
In all configurations, the application is decomposed into a DAG of tasks, each executing either a sequential Triton kernel or a parallel SYCL kernel via PoCL.  Figure~\ref{fig:kernels_hpccg_cpu_gran_multi} presents the iteration time as a function of task granularity (i.e., number of tasks) across three scenarios.

In the single runtime configuration (Single~RT, left), we use either the OpenMP or the OmpSs-2 runtime with sequential Triton kernels. This setup shows ideal scaling as performance improves steadily with the number of tasks, saturating around 256 tasks as the 112-core CPU becomes fully utilized. OmpSs-2 slightly outperforms OpenMP due to its more efficient task scheduling.

In the multiple runtime configuration (Multi.~RT, middle), the tasking runtime (OpenMP or OmpSs-2) is combined with PoCL, which executes parallel SYCL kernels. Here, the initial scaling trend reverses beyond 8--16 tasks due to oversubscription: each task spawns additional parallel threads via PoCL, resulting in thread contention and degraded performance. Compared to the single runtime case, peak performance is significantly lower.

In the multiple runtime with nOS-V configuration (Multi.\ w/~nOS-V,  right), we substitute the original OpenMP and PoCL runtimes with their nOS-V-enabled counterparts (\texttt{libompv} and \texttt{PoCLv}, respectively), allowing all runtimes to share a unified threading substrate. This change eliminates the oversubscription issue, allowing the application to scale well up to 64 tasks. While still slightly below the single runtime baseline, the performance is significantly improved over the multiple runtime setup without nOS-V, demonstrating the benefit of \emph{cooperative} runtimes.

\begin{figure}[htpb]
    \centering
    \includegraphics{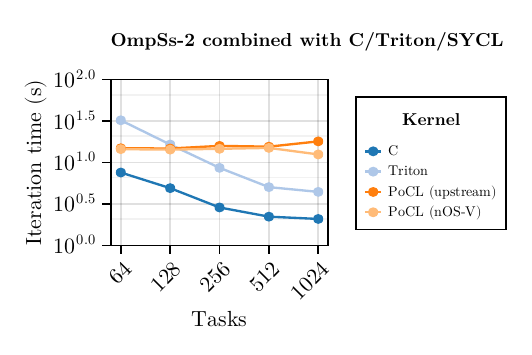}
    \caption{
        GPT-2 execution on CPU. Task-based execution with varying context length $T$.  Batch and channel dimensions remain fixed at 4 and 16 tasks, respectively.
    }
    \label{fig:gpt2_gran_splits_cpu_t}
\end{figure}

\paragraph{GPT-2 task-based execution}
We extend the task granularity analysis to the GPT-2 pre-training workload. Figure~\ref{fig:gpt2_gran_splits_cpu_t} shows the iteration time as we vary the context length $T \in \{1, 2, 4, 8, 16, 32\}$, while keeping the batch and channel dimensions fixed at 4 and 16 tasks, respectively. This results in a task count ranging from $4 \times 16 \times 1 = 64$ up to $2048$ tasks. The system has 112 cores, allowing us to observe both the underutilized and oversaturated regimes.

On the one hand, when using single-threaded kernels written in C or Triton, the application scales well up to approximately 1024 tasks, after which performance plateaus.  On the other hand, when each task launches a parallel SYCL kernel via PoCL, performance is significantly worse and largely flat across all task granularities regardless of whether PoCL is upstream or ported to nOS-V. This is due to excessive kernel-launch overhead from fine-grained tasks: even at the coarsest setting, there are already 64 tasks, each invoking a parallel kernel. The root bottleneck here is not runtime interference but rather the overhead of launching small parallel kernels, which masks any potential benefit from nOS-V.

These results highlight a common pitfall in heterogeneous multi-core applications: combining multiple runtime systems ---either explicitly through task-based models or implicitly via parallel libraries such as MKL--- can lead to severe performance degradation due to thread oversubscription and lack of coordination between runtimes. This effect is evident in both fork-join and task-based executions. By unifying all runtimes under a common threading and tasking substrate like nOS-V, these interferences can be largely mitigated, enabling more predictable and scalable performance. However, it remains the responsibility of the application developer to tune task granularity appropriately, as overly fine-grained tasks can still introduce severe inefficiencies even in the absence of runtime contention.

\section{Conclusions and future work}\label{sec:conclusions}
This work has demonstrated that combining the task-based data-flow model with Task-Aware libraries (TA-libs) offers a practical and flexible methodology for developing complex applications on heterogeneous systems. Our focus has been on how programmability, extensibility, and flexibility are enhanced by representing computations as DAGs and coordinating the use of multiple APIs, including SYCL, Triton, OpenMP offload, as well as vendor-specific libraries, within a unified task-based runtime. We validated this approach using two representative workloads: the GPT-2 pre-training phase, exemplifying a modern AI workload, and the HPCCG benchmark, illustrating a traditional HPC application. The task-aware abstraction enables seamless integration of diverse operations, even when fine-grained control over accelerators or communication is required, all while maintaining the advantages of dynamic scheduling and concurrency.

We have also shown that potential interference between different runtime systems can be minimized by leveraging a shared low-level infrastructure, such as the nOS-V threading and tasking library. Our experiments combining OmpSs-2, OpenMP, and SYCL via the P\ab{o}CL runtime demonstrate that interoperable and efficient executions are indeed possible without sacrificing programmability or performance portability.

The current evaluation has been limited to a single node and a single GPU device in order to isolate the core challenges of integrating multiple programming models. In future work, we plan to extend this methodology to distributed memory environments, leveraging task-aware communication libraries such as TAMPI. This would enable studies that combine accelerator-focused TA-libs (e.g., TACUDA and TASYCL) with communication TA-libs (e.g., TAMPI and TAGASPI), allowing unified orchestration of computation and communication across nodes and heterogeneous accelerators. Such an evaluation would also provide novel insights into inter-device and inter-node data movement.

\ab{
An additional direction for future research involves exploring finer-grained
task decomposition strategies that could benefit from kernel-level optimizations
such as CUDA Graphs/stream capture and Triton kernel fusion. Our current
methodology targets coarse-grained tasks that encapsulate substantial
computation, where kernel launch overhead is negligible. However, applications
with inherently fine-grained parallelism or those requiring numerous small
kernel invocations could benefit from mechanisms that reduce launch overhead and
enable kernel fusion opportunities. Investigating this regime would require
developing new benchmarks with fundamentally different task granularities and
analyzing the trade-offs between increased scheduling flexibility and reduced
per-kernel overhead. Such work would complement our current approach by
characterizing the full spectrum of task granularity options available within
unified task-based frameworks.
}

Finally, integrating TA-libs with transparent directory/cache (D/C) mechanisms found in modern task-based runtimes is a promising direction. This would combine the flexibility and explicit control of task-aware APIs with the automation and convenience of D/C-driven data management, further simplifying the development of scalable and portable applications for the next-generation of heterogeneous and distributed systems.

\section*{Code Availability}

The source code and evaluation scripts developed for this work, including contributions to GPT-2, HPCCG, nOS-V, and PoCL, are publicly available in a permanent repository at \url{https://doi.org/10.5281/zenodo.17534763}.

\section*{Acknowledgements}\label{sec:acknowledgements}

This work has received support from multiple funding sources. It is part of the ST4HPC project (PID2023-147979NB-C21), funded by MCIN/\allowbreak AEI/\allowbreak 10.13039/\allowbreak 5011000\allowbreak{}11033 and by FEDER, UE. This work is also promoted by the Barcelona Zettascale Laboratory, backed by the Ministry for Digital Transformation and of Public Services, within the framework of the Recovery, Transformation, and Resilience Plan - funded by the European Union - NextGenerationEU.
This work has received funding from the DARE SGA1 Project, from the European High-Performance Computing Joint Undertaking (JU) under Grant Agreement No 101202459 and from PCI2024-161687-3 Project funded by MICIU/\allowbreak{}AEI/\allowbreak{}10.13039/\allowbreak{}501100011033 and European's Union NextGenerationEU/PRTR.
Additional support was provided by a Ram\'on y Cajal fellowship (RYC2019-027592-I), funded by MCIN/\allowbreak AEI/\allowbreak 10.13039/\allowbreak 501100011033 and ESF/\allowbreak 10.13039/\allowbreak 501100004895, and by the Severo Ochoa Centre of Excellence programme (CEX2021-001148-S), also funded by MCIN/\allowbreak AEI. Finally, the Programming Models research group at BSC-UPC received support from the Departament de Recerca i Universitats de la Generalitat de Catalunya under grants 2021 SGR 01007 and 2025 STEP 00523.

\bibliographystyle{elsarticle-num-names}
\bibliography{bibliography.bib}

@article{Dongarra2016,
  author = {Jack Dongarra and Michael A Heroux and Piotr Luszczek},
  title ={High-performance conjugate-gradient benchmark: A new metric for ranking high-performance computing systems},
  journal = {The International Journal of High Performance Computing Applications},
  volume = {30},
  number = {1},
  pages = {3-10},
  year = {2016},
  doi = {10.1177/1094342015593158},
}

@article{Martinez2023-HLAM,
  author   = {Pedro J. Martinez-Ferrer and Tufan Arslan and Vicenç Beltran},
  title    = {Improving the performance of classical linear algebra iterative methods via hybrid parallelism},
  journal  = {Journal of Parallel and Distributed Computing},
  year     = {2023},
  volume   = {179},
  pages    = {104711},
  issn     = {0743-7315},
  doi      = {10.1016/j.jpdc.2023.04.012}
}

@article{Alvarez2023-taskiter,
  author   = {Álvarez, David and Beltran, Vicenç},
  title    = {Optimizing Iterative Data-Flow Scientific Applications Using Directed Cyclic Graphs},
  journal  = {IEEE Access},
  year     = {2023},
  volume   = {11},
  pages    = {51971-51984},
  issn     = {2169-3536},
  doi      = {10.1109/ACCESS.2023.3269902},
}

@misc{OpenAI2019-GPT2-pytorch,
  author       = {OpenAI},
  title        = {{GPT}-2 {PyTorch} implementation},
  howpublished = {https://github.com/openai/gpt-2/blob/master/src/model.py},
  year         = {2019},
  note         = {Accessed: 2025-05-01},
}

@misc{karpathy2024-tinyshakespeare,
  author       = {Andrej Karpathy},
  title        = {TinyShakespeare training dataset},
  howpublished = {https://raw.githubusercontent.com/karpathy/char-rnn/master/data/tinyshakespeare/input.txt},
  year         = {2024},
  note         = {Accessed: 2025-05-01},
}

@misc{Martinez2022-HLAM,
  title = {{HLAM}: Hybrid linear algebra methods},
  author = {Pedro J. Martinez-Ferrer},
  doi = {10.24433/CO.1193962.v1},
  year = 2022,
  date = {2022-09-07},
  version = {v1}
}

@book{Saad2003,
  doi = {10.1137/1.9780898718003},
  year = {2003},
  month = {jan},
  publisher = {Society for Industrial and Applied Mathematics},
  author = {Yousef Saad},
  title = {Iterative methods for sparse linear systems}
}

@techreport{Heroux2009,
  author={Michael A. Heroux and Douglas W. Douglas and Paul S. Crozier and James M. Willenbring and H. Carter Edwards and et al.},
  title={Improving performance via mini-applications},
  year={2009},
  number={SAND2009-5574},
  institution={Sandia National Laboratories},
  pages={1-40},
  url={https://mantevo.github.io/pdfs/MantevoOverview.pdf}
}

@Misc{Heroux2017,
  author       = {Michael A.~Heroux},
  title        = {High Performance Computing Conjugate Gradients ({HPCCG}): The original {Mantevo} miniapp},
  howpublished = {https://github.com/Mantevo/HPCCG},
  year         = {2017},
  note         = {Accessed: 2025-05-01},
}

@Misc{Karpathy2024-llm.c,
  author       = {Andrej Karpathy},
  title        = {{GPT}-2 {C} implementation},
  howpublished = {https://github.com/karpathy/llm.c},
  year         = {2024},
  note         = {Accessed: 2025-05-01},
}

@Misc{OpenAI2019-GPT2-pytorch-model,
  author       = {OpenAI},
  title        = {{GPT}-2 {PyTorch} model weights},
  howpublished = {https://huggingface.co/openai-community/gpt2/blob/main/pytorch\_model.bin},
  year         = {2021},
  note         = {Accessed: 2025-05-01},
}

@InProceedings{Sala2021-combining,
  author    = {Sala, Kevin and Macià, Sandra and Beltran, Vicenç},
  title     = {Combining One-Sided Communications with Task-Based Programming Models},
  booktitle = {2021 IEEE International Conference on Cluster Computing (CLUSTER)},
  year      = {2021},
  pages     = {528-541},
  month     = {Sep.},
  doi       = {10.1109/Cluster48925.2021.00024},
  issn      = {2168-9253},
}

@InProceedings{Sala2024-ALPI,
  author    = {Sala, Kevin and {\'A}lvarez, David and Pe{\~{n}}acoba, Ra{\'u}l and Arias Mallo, Rodrigo and Navarro, Antoni and Roca, Aleix and Beltran, Vicen{\c{c}}},
  title     = {{ALPI}: Enhancing Portability and Interoperability of Task-Aware Libraries},
  booktitle = {Asynchronous Many-Task Systems and Applications},
  year      = {2024},
  pages     = {142--153},
  isbn      = {978-3-031-61763-8},
}

@Article{Augonnet2011-StarPU,
  author     = {Augonnet, C\'{e}dric and Thibault, Samuel and Namyst, Raymond and Wacrenier, Pierre-Andr\'{e}},
  title      = {{StarPU}: a unified platform for task scheduling on heterogeneous multicore architectures},
  journal    = {Concurr. Comput.: Pract. Exper.},
  year       = {2011},
  volume     = {23},
  number     = {2},
  pages      = {187--198},
  month      = feb,
  issn       = {1532-0626},
  address    = {GBR},
  doi        = {10.1002/cpe.1631},
  issue_date = {February 2011},
  numpages   = {12},
  publisher  = {John Wiley and Sons Ltd.},
  url        = {https://doi.org/10.1002/cpe.1631},
}

@Article{Duran2011-OmpSs,
  author   = {Duran, Alejandro and Ayguad\'{e}, Eduard and Badia, Rosa M. and Labarta, Jes\'{u}s and Martinell, Luis and Martorell, Xavier and Planas, Judit},
  title    = {{OmpSs}: A PROPOSAL FOR PROGRAMMING HETEROGENEOUS MULTI-CORE ARCHITECTURES},
  journal  = {Parallel Processing Letters},
  year     = {2011},
  volume   = {21},
  number   = {02},
  pages    = {173-193},
  doi      = {10.1142/S0129626411000151},
}

@InProceedings{Gautier2007-KAAPI,
  author    = {Gautier, Thierry and Besseron, Xavier and Pigeon, Laurent},
  title     = {{KAAPI}: A thread scheduling runtime system for data flow computations on cluster of multi-processors},
  booktitle = {Proceedings of the 2007 International Workshop on Parallel Symbolic Computation},
  year      = {2007},
  series    = {PASCO '07},
  pages     = {15--23},
  address   = {New York, NY, USA},
  publisher = {Association for Computing Machinery},
  doi       = {10.1145/1278177.1278182},
  isbn      = {9781595937414},
  location  = {London, Ontario, Canada},
  numpages  = {9},
}

@InProceedings{Kim2021-IRIS,
  author    = {Kim, Jungwon and Lee, Seyong and Johnston, Beau and Vetter, Jeffrey S.},
  title     = {{IRIS}: A Portable Runtime System Exploiting Multiple Heterogeneous Programming Systems},
  booktitle = {2021 IEEE High Performance Extreme Computing Conference (HPEC)},
  year      = {2021},
  pages     = {1-8},
  month     = {Sep.},
  doi       = {10.1109/HPEC49654.2021.9622873},
  issn      = {2643-1971},
}

@InProceedings{Memeti2017-benchmarking,
  author    = {Memeti, Suejb and Li, Lu and Pllana, Sabri and Ko\l{}odziej, Joanna and Kessler, Christoph},
  title     = {Benchmarking {OpenCL}, {OpenACC}, {OpenMP}, and {CUDA}: Programming Productivity, Performance, and Energy Consumption},
  booktitle = {Proceedings of the 2017 Workshop on Adaptive Resource Management and Scheduling for Cloud Computing},
  year      = {2017},
  series    = {ARMS-CC '17},
  pages     = {1--6},
  address   = {New York, NY, USA},
  publisher = {Association for Computing Machinery},
  doi       = {10.1145/3110355.3110356},
  isbn      = {9781450351164},
  location  = {Washington, DC, USA},
  numpages  = {6}
}

@InProceedings{Reguly2023-evaluating,
  author    = {Reguly, Istvan Z.},
  title     = {Evaluating the performance portability of {SYCL} across {CPUs} and {GPUs} on bandwidth-bound applications},
  booktitle = {Proceedings of the SC '23 Workshops of the International Conference on High Performance Computing, Network, Storage, and Analysis},
  year      = {2023},
  series    = {SC-W '23},
  pages     = {1038--1047},
  address   = {New York, NY, USA},
  publisher = {Association for Computing Machinery},
  doi       = {10.1145/3624062.3624180},
  isbn      = {9798400707858},
  location  = {Denver, CO, USA},
  numpages  = {10}
}

@InProceedings{Korakitis2022-towards,
  author    = {Korakitis, Orestis and De Gonzalo, Simon Garcia and Guidotti, Nicolas and Barreto, Jo\~{a}o Pedro and Monteiro, Jos\'{e} C. and Pe\~{n}a, Antonio J.},
  title     = {Towards {OmpSs-2} and {OpenACC} interoperation},
  booktitle = {Proceedings of the 27th ACM SIGPLAN Symposium on Principles and Practice of Parallel Programming},
  year      = {2022},
  series    = {PPoPP '22},
  pages     = {433--434},
  address   = {New York, NY, USA},
  publisher = {Association for Computing Machinery},
  doi       = {10.1145/3503221.3508401},
  isbn      = {9781450392044},
  location  = {Seoul, Republic of Korea},
  numpages  = {2}
}

@Article{Dagum1998-OpenMP,
  author   = {Dagum, L. and Menon, R.},
  title    = {{OpenMP}: an industry standard {API} for shared-memory programming},
  journal  = {IEEE Computational Science and Engineering},
  year     = {1998},
  volume   = {5},
  number   = {1},
  pages    = {46-55},
  month    = jan,
  issn     = {1558-190X},
  doi      = {10.1109/99.660313},
}

@InProceedings{Perez2017-OmpSs-2,
  author    = {Perez, Josep M. and Beltran, Vicenç and Labarta, Jesus and Ayguadé, Eduard},
  title     = {Improving the Integration of Task Nesting and Dependencies in {OpenMP}},
  booktitle = {2017 IEEE International Parallel and Distributed Processing Symposium (IPDPS)},
  year      = {2017},
  pages     = {809-818},
  doi       = {10.1109/IPDPS.2017.69},
}

@Article{Bosilca2013-PaRSEC,
  author   = {Bosilca, George and Bouteiller, Aurelien and Danalis, Anthony and Faverge, Mathieu and Herault, Thomas and Dongarra, Jack J.},
  title    = {{PaRSEC}: Exploiting Heterogeneity to Enhance Scalability},
  journal  = {Computing in Science \& Engineering},
  year     = {2013},
  volume   = {15},
  number   = {6},
  pages    = {36-45},
  month    = {Nov},
  issn     = {1558-366X},
  doi      = {10.1109/MCSE.2013.98},
}

@misc{Heroux2017-HPCCG,
  author = {Michael A. Heroux},
  title  = {High Performance Computing Conjugate Gradients ({HPCCG}): The original {Mantevo} miniapp},
  year   = {2017},
  date   = {2017-07-04},
  url    = {https://github.com/Mantevo/HPCCG},
}

@Article{Radford2019-GPT2,
  author   = {Radford, Alec and Wu, Jeffrey and Child, Rewon and Luan, David and Amodei, Dario and Sutskever, Ilya and others},
  title    = {Language models are unsupervised multitask learners},
  year     = {2019},
  url      = {https://cdn.openai.com/better-language-models/language_models_are_unsupervised_multitask_learners.pdf},
}

@Article{Radford2018-GPT,
  author   = {Alec Radford and Karthik Narasimhan and Tim Salimans and Ilya Sutskever},
  title    = {Improving language understanding by generative pre-training},
  year     = {2018},
  url      = {https://cdn.openai.com/research-covers/language-unsupervised/language_understanding_paper.pdf},
}

@Article{Wei2022-emergentLLMs,
  author   = {Jason Wei and Yi Tay and Rishi Bommasani and Colin Raffel and Barret Zoph and Sebastian Borgeaud and Dani Yogatama and Maarten Bosma and Denny Zhou and Donald Metzler and Ed H. Chi and Tatsunori Hashimoto and Oriol Vinyals and Percy Liang and Jeff Dean and William Fedus},
  title    = {Emergent Abilities of Large Language Models},
  journal  = {Transactions on Machine Learning Research},
  year     = {2022},
  issn     = {2835-8856},
  url      = {https://openreview.net/forum?id=yzkSU5zdwD},
}

@InProceedings{Vaswani2017-attTransformer,
  author    = {Vaswani, Ashish and Shazeer, Noam and Parmar, Niki and Uszkoreit, Jakob and Jones, Llion and Gomez, Aidan N. and Kaiser, \L{}ukasz and Polosukhin, Illia},
  title     = {Attention is all you need},
  booktitle = {Proceedings of the 31st International Conference on Neural Information Processing Systems},
  year      = {2017},
  series    = {NIPS'17},
  pages     = {6000--6010},
  address   = {Red Hook, NY, USA},
  publisher = {Curran Associates Inc.},
  isbn      = {9781510860964},
  location  = {Long Beach, California, USA},
  numpages  = {11},
  doi       = {10.5555/3295222.3295349}
}

@Misc{ROCm,
  author       = {AMD},
  title        = {{AMD} {ROCm} documentation},
  howpublished = {https://rocm.docs.amd.com/en/latest/},
  year         = {2025},
  note         = {Accessed: 2025-05-01},
}

@Misc{OpenCL,
  author       = {Khronos~Group},
  title        = {{OpenCL} for Parallel Programming of Heterogeneous Systems},
  howpublished = {https://www.khronos.org/opencl/},
  year         = {2025},
  note         = {Accessed: 2025-05-01},
}

@Misc{Triton,
  author       = {Triton},
  title        = {Triton documentation},
  howpublished = {https://triton-lang.org/main/index.html},
  year         = {2025},
  note         = {Accessed: 2025-05-01},
}

@Misc{SYCL,
  author       = {Khronos~Group},
  title        = {SYCL: C++ Programming for Heterogeneous Parallel Computing},
  howpublished = {https://www.khronos.org/sycl/},
  year         = {2025},
  note         = {Accessed: 2025-05-01},
}

@Misc{OpenACC,
  author       = {OpenACC-Standard.org},
  title        = {The {OpenACC}{\textregistered} Application Programming Interface},
  howpublished = {https://www.openacc.org/sites/default/files/inline-images/Specification/OpenACC-3.3-final.pdf},
  year         = {2022},
  note         = {Accessed: 2025-05-01},
}

@Misc{CUDA,
  author       = {NVIDIA},
  title        = {{CUDA} Toolkit Documentation},
  howpublished = {https://docs.nvidia.com/cuda/},
  year         = {2025},
  note         = {Accessed: 2025-05-01},
}

@Article{Yi2019,
author={Yi, Gangman and Loia, Vincenzo},
title={High-performance computing systems and applications for AI},
journal={The Journal of Supercomputing},
year={2019},
month={Aug},
day={01},
volume={75},
number={8},
pages={4248-4251},
issn={1573-0484},
doi={10.1007/s11227-019-02937-z},
url={https://doi.org/10.1007/s11227-019-02937-z}
}

@Article{Huerta2020,
author={Huerta, E. A.
and Khan, Asad
and Davis, Edward
and Bushell, Colleen
and Gropp, William D.
and Katz, Daniel S.
and Kindratenko, Volodymyr
and Koric, Seid
and Kramer, William T. C.
and McGinty, Brendan
and McHenry, Kenton
and Saxton, Aaron},
title={Convergence of artificial intelligence and high performance computing on NSF-supported cyberinfrastructure},
journal={Journal of Big Data},
year={2020},
month={Oct},
day={16},
volume={7},
number={1},
pages={88},
issn={2196-1115},
doi={10.1186/s40537-020-00361-2},
url={https://doi.org/10.1186/s40537-020-00361-2}
}

@InProceedings{Duran2008,
author="Duran, Alejandro
and Perez, Josep M.
and Ayguad{\'e}, Eduard
and Badia, Rosa M.
and Labarta, Jesus",
editor="Eigenmann, Rudolf
and de Supinski, Bronis R.",
title="Extending the OpenMP Tasking Model to Allow Dependent Tasks",
booktitle="OpenMP in a New Era of Parallelism",
year="2008",
publisher="Springer Berlin Heidelberg",
address="Berlin, Heidelberg",
pages="111--122",
isbn="978-3-540-79561-2"
}

@Article{Duran2009,
author={Duran, Alejandro
and Ferrer, Roger
and Ayguad{\'e}, Eduard
and Badia, Rosa M.
and Labarta, Jesus},
title={A Proposal to Extend the OpenMP Tasking Model with Dependent Tasks},
journal={International Journal of Parallel Programming},
year={2009},
month={Jun},
day={01},
volume={37},
number={3},
pages={292-305},
issn={1573-7640},
doi={10.1007/s10766-009-0101-1},
url={https://doi.org/10.1007/s10766-009-0101-1}
}

@INPROCEEDINGS{Converse,
  author={Kale, L.V. and Bhandarkar, M. and Jagathesan, N. and Krishnan, S. and Yelon, J.},
  booktitle={Proceedings of International Conference on Parallel Processing}, 
  title={Converse: an interoperable framework for parallel programming}, 
  year={1996},
  volume={},
  number={},
  pages={212-217},
  doi={10.1109/IPPS.1996.508060}}

@INPROCEEDINGS{MetaChaos,
  author={Edjlali, G. and Sussman, A. and Saltz, J.},
  booktitle={Proceedings 11th International Parallel Processing Symposium}, 
  title={Interoperability of data parallel runtime libraries}, 
  year={1997},
  volume={},
  number={},
  pages={451-459},
  doi={10.1109/IPPS.1997.580940}}

@inproceedings{HeteroBench,
author = {Tian, Hongzheng and Mishra, Alok and Chen, Zhiheng and Hong Enriquez, Rolando P. and Milojicic, Dejan and Frachtenberg, Eitan and Huang, Sitao},
title = {HeteroBench: Multi-kernel Benchmarks for Heterogeneous Systems},
year = {2025},
isbn = {9798400710735},
publisher = {Association for Computing Machinery},
address = {New York, NY, USA},
url = {https://doi.org/10.1145/3676151.3719366},
doi = {10.1145/3676151.3719366},
booktitle = {Proceedings of the 16th ACM/SPEC International Conference on Performance Engineering},
pages = {320--333},
numpages = {14},
location = {Toronto ON, Canada},
series = {ICPE '25}
}

@INPROCEEDINGS{Roberts2018,
  author={Roberts, Steve and Ramanna, Pradeep and Walthour, John},
  booktitle={2018 IEEE High Performance extreme Computing Conference (HPEC)}, 
  title={AC922 Data Movement for CORAL}, 
  year={2018},
  volume={},
  number={},
  pages={1-5},
  doi={10.1109/HPEC.2018.8547707}}

@inproceedings{GraceHopper,
author = {Schieffer, Gabin and Wahlgren, Jacob and Ren, Jie and Faj, Jennifer and Peng, Ivy},
title = {Harnessing Integrated CPU-GPU System Memory for HPC: a first look into Grace Hopper},
year = {2024},
isbn = {9798400717932},
publisher = {Association for Computing Machinery},
address = {New York, NY, USA},
url = {https://doi.org/10.1145/3673038.3673110},
doi = {10.1145/3673038.3673110},
booktitle = {Proceedings of the 53rd International Conference on Parallel Processing},
pages = {199--209},
numpages = {11},
location = {Gotland, Sweden},
series = {ICPP '24}
}

@article{jaaskelainen_pocl_2015,
	title = {pocl: A Performance-Portable {OpenCL} Implementation},
	volume = {43},
	year = {2015},
	issn = {1573-7640},
	url = {https://doi.org/10.1007/s10766-014-0320-y},
	doi = {10.1007/s10766-014-0320-y},
	shorttitle = {pocl},
	pages = {752--785},
	number = {5},
	journaltitle = {International Journal of Parallel Programming},
	shortjournal = {Int J Parallel Prog},
	author = {Jääskeläinen, Pekka and de La Lama, Carlos Sánchez and Schnetter, Erik and Raiskila, Kalle and Takala, Jarmo and Berg, Heikki},
	urldate = {2025-10-31},
	date = {2015-10-01},
	langid = {english},
}

@INPROCEEDINGS{alvareznosv,
  author={Álvarez, David and Sala, Kevin and Beltran, Vicenç},
  booktitle={2024 IEEE International Parallel and Distributed Processing Symposium (IPDPS)},
  title={nOS-V: Co-Executing HPC Applications Using System-Wide Task Scheduling},
  year={2024},
  volume={},
  number={},
  pages={312-324},
  doi={10.1109/IPDPS57955.2024.00035}
}

@article{navarro_synergizing_2025,
  title = {Synergizing {OpenMP} Paradigms Through Free Agents and {nOS}-V},
  volume = {6},
  year = {2025},
  issn = {2661-8907},
  url = {https://doi.org/10.1007/s42979-025-04406-2},
  doi = {10.1007/s42979-025-04406-2},
  pages = {869},
  number = {7},
  journaltitle = {{SN} Computer Science},
  shortjournal = {{SN} {COMPUT}. {SCI}.},
  author = {Navarro, Antoni and Peñacoba, Raúl and Arcila, Vincent A. and Arias, Rodrigo and Álvarez, David and Beltran, Vicenç},
  urldate = {2025-11-04},
  date = {2025-10-03},
  langid = {english},
}

@InProceedings{Dao2022-FlashAttention,
  author    = {Dao, Tri and Fu, Daniel Y. and Ermon, Stefano and Rudra, Atri and R\'{e}, Christopher},
  title     = {{FlashAttention}: fast and memory-efficient exact attention with {IO}-awareness},
  booktitle = {Proceedings of the 36th International Conference on Neural Information Processing Systems},
  year      = {2022},
  series    = {NIPS '22},
  address   = {Red Hook, NY, USA},
  publisher = {Curran Associates Inc.},
  articleno = {1189},
  doi       = {10.5555/3600270.3601459},
  isbn      = {9781713871088},
  location  = {New Orleans, LA, USA},
  numpages  = {16},
  url       = {https://dl.acm.org/doi/10.5555/3600270.3601459},
}

@online{cudatoolkit_launchhostfunc,
	title = {CUDA Toolkit Documentation --- cudaLaunchHostFunc},
	url = {https://docs.nvidia.com/cuda/cuda-runtime-api/group__CUDART__EXECUTION.html#group__CUDART__EXECUTION_1g05841eaa5f90f27124241baafb3e856f},
	type = {{cppModule}},
	urldate = {2025-11-07},
  year = {2025},
	langid = {english},
}

\appendix
\ab{
\section{Mixed composition}%
\label{sec:appendix:mixed}

%
\begin{table}[htb]
    \centering
    \caption{Programming model used for each kernel in HPCCG Mixed version.}%
    \label{tab:mixed_hpccg}
    \begin{tabular}{lcc}
        \toprule
                            & \multicolumn{2}{c}{Version} \\
                              \cmidrule(lr){2-3}
        Kernel              & CPU                          & GPU    \\
        \midrule
        SpMV*               & SYCL                         & Triton \\
        axpby               & SYCL                         & SYCL   \\
        axpy                & SYCL                         & Triton \\
        copy                & OpenMP                       & Triton \\
        dot                 & Triton                       & Triton \\
        reductions          & OpenMP                       & OpenMP \\
        \bottomrule
        \multicolumn{3}{p{15em}}{
            \footnotesize
            *This kernel was replaced by a vendor implementation (oneMKL / cuBLAS) on configurations that
            used vendor libraries.
        }
    \end{tabular}
\end{table}

\begin{table}[htb]
    \centering
    \caption{Programming model used for each kernel in GPT-2 Mixed version.}%
    \label{tab:mixed_gpt2} 
    \begin{tabular}{lcc}
        \toprule
                                          & \multicolumn{2}{c}{GPT-2 pass} \\
                          \cmidrule(lr){2-3}
        Kernel                            & Forward                         & Backward \\
        \midrule
        Attention                         & Triton                          & Triton   \\
        Crossentropy                      & SYCL                            & OpenMP   \\
        Encoder                           & SYCL                            & OpenMP   \\
        GELU                              & SYCL                            & OpenMP   \\
        Layernorm                         & Triton                          & Triton   \\
        Matmul*                           & Triton                          & --       \\
        Matmul (input)*                   & --                              & SYCL     \\
        Matmul (params)                   & --                              & Triton   \\
        Residual                          & SYCL                            & OpenMP   \\
        Softmax                           & Triton                          & OpenMP   \\
        \addlinespace[2ex]
        Update Step                       & \multicolumn{2}{c}{SYCL}       \\
        \bottomrule
        \multicolumn{3}{p{18em}}{
            \footnotesize
            *These kernels were replaced by vendor implementations (oneMKL / cuBLAS) on configurations that
            used vendor libraries.
        }
    \end{tabular}
\end{table}


}





\end{document}